# Effect of extended defects on phonon confinement in polycrystalline Si and Ge films

Larisa V. Arapkina[1], Kirill V. Chizh[1], Oleg V. Uvarov[1], Valery V. Voronov[1], Vladimir P. Dubkov[1], Mikhail S. Storozhevykh[1], Maksim V. Poliakov[2], Lidiya S. Volkova[2], Polina A. Edelbekova[2], Alexey A. Klimenko[2], Alexander A. Dudin[2] and Vladimir A. Yuryev[1]

[1] Prokhorov General Physics Institute of the Russian Academy of Sciences, 38 Vavilov Street, 119991 Moscow, Russia

[2] Institute of Nanotechnology of Microelectronics of the Russian Academy of Sciences, 32A Leninsky Prospekt, 119991 Moscow, Russia

Corresponding author:

Larisa V. Arapkina     arapkina@kapella.gpi.ru

**Abstract**

We present Raman spectroscopy of the polycrystalline Si and Ge films deposited by molecular beam deposition on a dielectric substrate. The Raman study has been made using lasers with different wavelengths. Structural properties of the poly-films have been studied by XRD and TEM. The Raman spectra are characterized by appearance of the additional wide peaks around 500 cm$^{-1}$ and 290 cm$^{-1}$ in the main vibrational bands of TO(c-Si) and TO(c-Ge) phonons, respectively. It is shown that these peaks correspond to scattering in grain boundary area. For the poly–Si films, both a downward shift and an asymmetrical broadening of the vibrational band of TO(c-Si) near 520 cm$^{-1}$ are observed, whereas there is only a symmetric broadening in the spectra of poly-Ge. The Raman line shape has been modeled within the framework of the phonon confinement theory taking into account the sizes of coherent scattering domains obtained using XRD. The model includes a symmetrical band broadening observed in polycrystalline films. It is shown that confinement of phonon propagation might be in the poly–Si films. The phonon dispersion and the density of phonon states have been simulated using density functional theory. It has been found that phonon confinement relates to grain boundaries rather than other extended defects such as twins (multiple twins, twin boundaries), the appearance of which does not lead to significant changes in phonon dispersion and density of phonon states.

# 1. Introduction

The structural and optical properties of the polycrystalline Si (*poly*–Si) and Ge (*poly*–Ge) films grown by molecular beam epitaxy (MBE) have been investigated in the present work. In the Raman spectra of the *poly*–Si, a downward shift and an asymmetrical broadening of the TO-phonon band at 521 cm$^{-1}$ have been observed. We have studied the possibility of the application of the phonon confinement model (PCM) to polycrystalline Si and Ge films bearing in mind that extended defects (grain boundaries, twins and etc.) might be considered as a reason of the phonon confinement and selection rule relaxation. PCM is a phenomenological model proposed by Richter [1] for description of Raman spectra of microcrystalline Si grown by the plasma transport method. It was based on the statement about the relaxation of the wave vector selection in small confinement domains in nanoscale structures. According to it, the effect of the phonon confinement is observed in spherical nanocrystals with sizes not exceeding ~10 nm. Then, this approach has been developed by Campbell and Fauchet [2] for nanocrystals with different shape (nanowires and thin layers) and despite of the application of an even stronger phonon confinement, the changes in the Raman curve are observed only for nanocrystals with size up to 3 nm. PCM has got an expansion in the works [3–11] and has been successfully applied in the description of Raman experimental data for the structures with quantum dots and nanowires [6,7,9,12–18] and quantum wells [19]. This model is particularly good at processing experimental data of samples with known nanocrystal sizes and their size distribution, and the understandable influence of temperature and elastic stresses. One of the main model concepts defining the Raman line shape is the confinement envelope function $W(q)$ determining the phonon localization on the nanocrystal boundary and reducing the phonon amplitude outside the nanocrystal. A frequently used function is Gaussian $W(q) = exp\{-(\alpha r/L)^2\}$ as, e.g., in [14] (sometimes authors use a different type of notation in relation to the parameter α). For Si and Ge nanocrystals a different value of the parameter α is applied, which determines the length scale for phonon confinement, so how quickly the phonon amplitude decreases towards the nanocrystal boundary [1,2,5,11,14,19–25]. The values of the parameter α proposed by Richter [1] and Campbell [2] might be considered as the extreme ones. The equation $W(q) = \exp(-4r^2/L^2)$ proposed by Richter corresponds to "soft confinement" on the boundary. The equation $W(q) = \exp(-8\pi^2 r^2/L^2)$ suggested by Campbell and Fauchet describes a case of "rigid" phonon localization into the small central part of a nanocrystal. When considering the phonon localization conditions at the nanocrystal boundary, it is necessary to take into account the matrix, inside which nanocrystal is formed, the presence of an oxide shell or an interface with material having different composition or structure. Here, the presence of overlap of the phonon dispersion curves of the nanocrystal and matrix (shell, layer) have the great impact on phonons propagating. If an overlap takes place then phonons might propagate into the neighboring layer and phonon confinement is weak as, for example, in some $A^3B^5$ heterostructures [19]. The samples are polycrystalline in the present work, and, in general, do not contain domains with dimension suitable for PCM consideration because they are composed of the larger columnar grains than necessary to observe the phonon confinement in Raman spectra [1]. We have tried to apply the PCM considering phonon propagation under the condition of extended defects situated in grains and grain boundaries. We combined the structural researching by the X-ray diffraction (XRD) and transmission electron microscopy (TEM) with the Raman study using the lasers with different wavelengths. The *poly*–Si (*poly*–Ge) films were grown at different growth temperatures. Raman spectrum were simulated within the framework of PCM taking into account the broadening of the Raman spectrum of a polycrystalline film. We used density functional theory (DFT) to model phonon spectra for the structures containing several extended defects.

## 2. Sample preparation and experimental methods

### 2.1. Preparation of samples

The experimental samples were prepared by depositing *poly*–Si and *poly*–Ge layers on $Si_3N_4/SiO_2/Si$ (001) dielectric substrates. A detailed description of the used techniques of the sample production including preliminary chemical treatment of $Si_3N_4/SiO_2/Si$ (001) substrates can be found in our previous works [26,27]. Si and Ge films were deposited in ultra-high vacuum condition in an EVA 32 molecular beam epitaxy (MBE) chamber (Riber). After wet chemical treatment, the substrates were annealed at 600°C for 6 hours at pressure of less than $5 \times 10^{-9}$ Torr in the preliminary annealing chamber. The MBE growth chamber was evacuated down to about $3 \times 10^{-11}$ Torr before the processes; the pressure did not exceed $5 \times 10^{-9}$ Torr during the Si or Ge deposition. The thicknesses of the Si layers were 2, 50 and 200 nm and those of Ge films were 2, 50, 100 and 200 nm. The deposition rate was ~ 0.25 Å/s. The film growth temperature was in the range from 400 to 650 °C. The growth temperature control was carried out using thermocouples graduated against an IMPAC IS12-Si infrared pyrometer (LumaSense Technologies). The residual atmosphere composition in the MBE growth chamber was monitored using an RGA-200 residual gas analyzer (Stanford Research Systems). Other process details might be found in [26,27].

### 2.2. Experimental methods

The main experimental methods were RHEED, XRD, TEM and Raman spectroscopy.

RHEED images were acquired *in situ* during the deposition processes and after sample cooling to the room temperature. RHEED patterns were processed as those of polycrystalline silicon and germanium surfaces. A RH20 reflection RHEED tool (Staib Instruments) was used.

Raman spectra were registered using an inVia Qontor spectrometer (Renishaw) in a backscattering geometry. The spectra recording was realized using vertically polarized laser excitation at the wavelengths of 405, 532, 633 and 785 nm with the emission power of 15, 10, 8.5 and 5.4 mW, respectively, which were selected to avoid heating the sample during measurement. A diffraction grating with 1200 gr/mm was used for 785-nm and 633-nm lasers and that with 2400 gr/mm was applied for 405-nm and 532-nm ones; a 50× objective was employed in all the experiments. Using lasers with several wavelengths enabled obtaining the information from different depths in the sample. The depth of laser radiation penetration in the polycrystalline films have been estimated using the optical constants of *c*-Si (*c*-Ge) and amorphous Si (*α*-Si) and Ge (*α*-Ge) [28,29]. The measurements were carried out at room temperature. Voigt, Pseudo-Voigt and Gauss functions were used for the peak analysis of Raman bands; *c*-Si, *c*-Ge and $Si_3N_4/SiO_2$/Si (001) substrates were used as reference samples.

X-ray diffraction measurements of the *poly*–Si films were carried out with a D2 PHASER diffractometer (Bruker) in the ranges of diffraction angles $2\theta$ = 10 to 60° with a step along the angle $\Delta(2\theta) = 0.05°$ and a 100 μm wide collimator slit. The exposure at each point was 15 s. The radiation source was Cu $K_\alpha$ or Cu $K_\beta$. The samples with *poly*–Ge films were measured using D8 Advance diffractometer (Bruker) in the range of diffraction angle $2\theta$ = 10 to 120° with a step along the angle $\Delta(2\theta) = 0.02$ or 0.005°. The radiation source was Cu $K_\alpha$. The coherent scattering domain size (*D*) was determined taking into account the instrumental broadening, which was found using reference Si and Ge samples. The influence of micro-stresses in the polycrystalline films on peak width was not considered. Thus, the calculated value of *D* is considered as the smallest coherent scattering domain. XRD spectra were fitted by Gaussian function. The ICDD PDF-2 database was used for peak identification.

A Libra 200 FE HR TEM instrument (Carl Zeiss) was used for transmission electron microscopy. Samples for HR TEM were cut and then thinned in $Ar^+$ plasma using a Model 1010 ion mill (E.A.Fischione Instruments) at an ion accelerating voltage of 4 kV. Some samples were examined using a JEM 2100 Plus (JEOL) transmission electron microscope operating at 200 kV; these samples were prepared with a focused beam of Ga+ ion using a FEI Helios G4 CX scanning electron microscope (Thermo Fisher Scientific). In this case, the samples were coated with a protective layer of Pt, which is visible in the TEM images.

## 3. Results and Discussion
### 3.1. Structure of the polycrystalline Si and Ge layers
### 3.1.1. Reflected high-energy electron diffraction

Fig.1 and Fig.2 present RHEED images of the surface of the 2 nm, 50 nm and 200 nm thick Si and Ge films grown at 500°C, respectively. All of them consist of diffuse Debye rings confirming that the Si and Ge films are polycrystalline. The radii of the rings correspond to the reciprocal interplanar distances ($1/d_{hkl}$) in the lattices of silicon or germanium [30], which have the $Fd\bar{3}m$ (227) space group. In Fig.1(a) and Fig.2(a), the RHEED images are related to the thinnest samples comprising 2 nm thick films of Si and Ge, respectively. All rings are seen to be continuous, which means that crystalline grains forming the films are randomly oriented. In the pattern of the Si film (Fig.1(a)), broad intense reflexes are observed, so that neighboring rings overlap each other and cannot be resolved (see, e.g., the (422) and (333) pair). For the Ge film (Fig.2(a)), the rings are more intense and narrower than for the Si layer and adjacent reflexes are quite resolvable. The comparison of the 2 nm thick Si and Ge samples shows that the Ge films consist of larger grains at the initial growth stage than the Si layers do. It should be expected that *poly*–Ge film is entirely composed of larger grains than the *poly*–Si one. With an increase in the thickness of the Si film (Fig.1(b,c)), the rings become narrower and discontinuous ones appear corresponding to formation of coarser grains with a preferential growth orientation in the near surface domains. Similar scenario is observed for the Ge deposition (Fig.2(b,c)). The broken rings emerge with the increasing Ge film thickness as well; there is some narrowing of the lines but, as in Fig.1(b,c), neighboring rings could not be resolved either because of the strong background scattering.

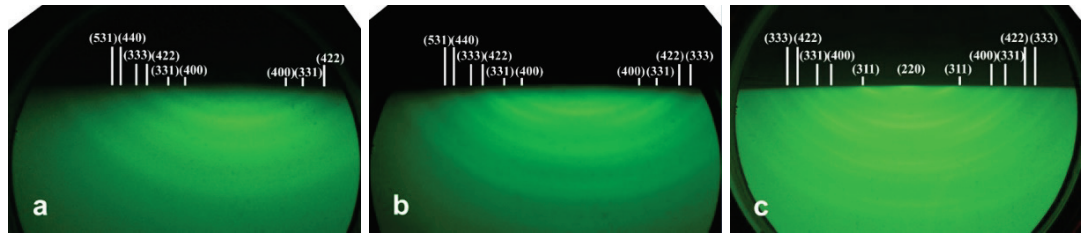

Fig.1. RHEED patterns of the surface of a Si layer grown at 500°C; the film thicknesses are (a) 2, (b) 50 and (c) 200 nm.

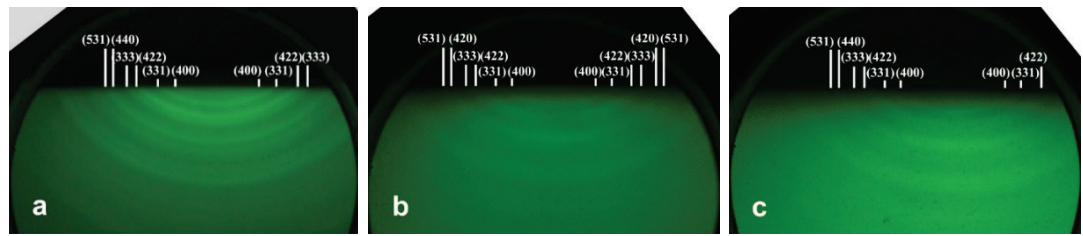

Fig.2. RHEED patterns of the surface of a Ge layer grown at 500°C; the film thicknesses are (a) 2, (b) 50 and (c) 200 nm.

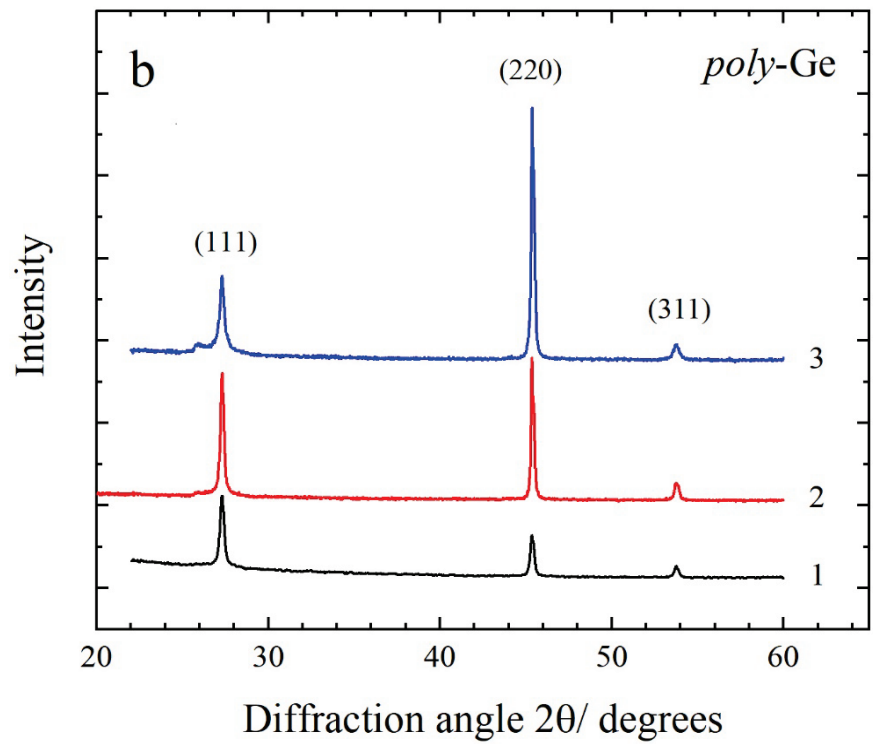

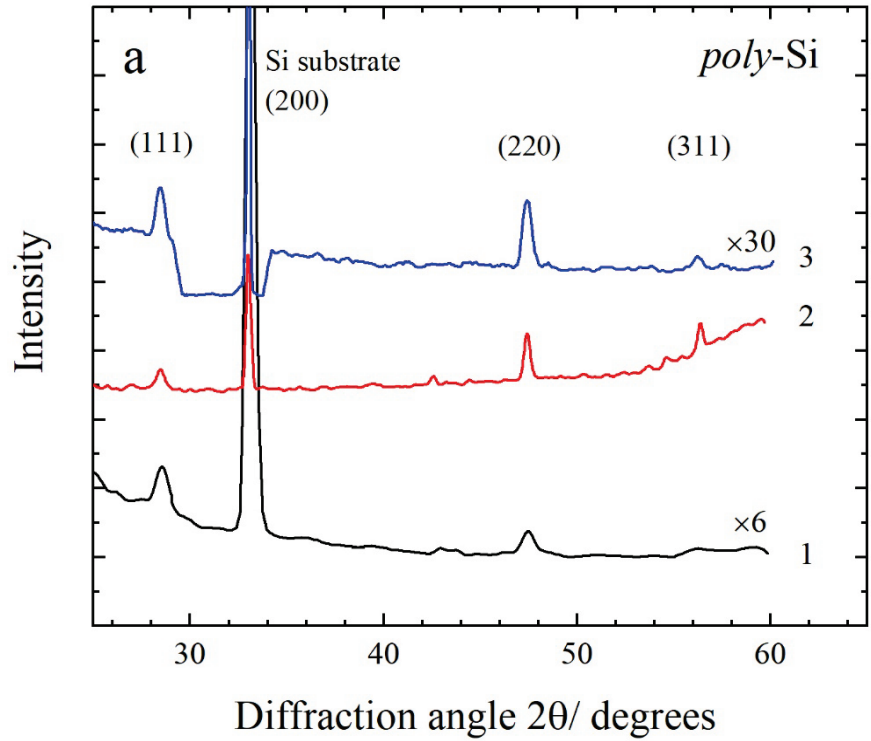

Fig.3 X-ray diffraction patterns of the *poly*-Si (a) and *poly*-Ge (b) films; in panel (a), the thicknesses and deposition temperatures of the films are as follows: (1) 50 nm, 500°C, (2) 200 nm, 500°C, (3) 200 nm, 650°C; in panel (b), they are (1) 50 nm, 500°C, (2) 200 nm, 500°C, (3) 200 nm, 400°C; the curves are shifted for clarity.

### 3.1.2. X-ray diffraction

The XRD patterns measured in the *poly*–Si films are presented in Fig.3(a). The main diffraction peaks of this pattern are assigned to reflections from the (111), (200), (220) and (331) planes, that corresponds to the crystal orientations (111), (100), (110) and (331), respectively, with the strongest one being related to the reflection from the (200) plane of the Si (001) substrate, whereas the others being due to the reflections from the crystal planes of the *poly*–Si films. At the growth temperature of 500°C, with increasing film thickness, the preferential orientation of film grains changes from the (111) and (110) (curve 1) to (110) and (331) (curve 2). For films 200 nm thick, the intensity of the (111) diffraction peak increases with increasing film deposition temperature (curve 2 and curve 3). Thus, there is no single preferential orientation of crystallites in the poly–Si film and hence one should expect a large number of incoherent grain boundaries (GB) there, especially in regions close the interface with the substrate. Such GBs usually result in distortion of the crystal lattice. The coherent scattering domains with the smallest sizes calculated using the Scherrer equation are about 26 nm for the 50 nm and 200 nm thick films grown at 500°C and about 30 nm for the 200 nm thick ones grown at 650°C. These value are close for the (111) and (110) crystal orientations.

Fig.3(b) shows the XRD patterns of the *poly*-Ge films that have been obtained with 2–2.5° tilt of samples to reduce the influence of the X-ray tube signals. The main diffraction peaks correspond to (111), (110) and (331) crystal orientations. In the samples grown at 500°C, with increasing *poly*–Ge film thickness from 50 to 200 nm, the ratio of the intensities of the (111) and (220) diffraction peaks changes and the (220) one grows up (curve 1 and curve 2). Any preferential orientation of grains is absent in these samples. For the growth temperature of 400°C (curve 3), the diffraction peak (220) predominates. In the poly–Ge films, a noticeable difference in the sizes of coherent scattering domains is observed for [110] and [111] growth directions, since the growth rate is the highest in the [110] direction. For the samples grown at 500°C, a size ratio increases 1.5 to 2.2 times, as a thickness changes from 50 to 200 nm. The smallest value of $D$ is observed for the (111) one; it is about 54 nm for the 50 and 100 nm thick *poly*–Ge films and about 83 nm for the 200 nm thick ones grown at 500°C. In the 200-nm film deposited at 400°C, it is about 50 nm. The spectra of the *poly*–Ge films (curve 2 and curve 3 in Fig.3(b)) show a weak feature in the form of an additional diffraction peak placed next to the (111) one; this feature appears owing to the increase in the interplanar distance in the [111] direction in some domains [31]. We suppose that these domains might be formed at GBs.

### 3.1.3. Transmission electron microscopy

The structure of the *poly*–Si and *poly*–Ge layers have been explored using TEM.

STEM images of the 200 nm thick Si layers deposited at 450°C and 500°C are presented in Fig.4.[1] In the *poly*–Si film grown at 450°C (Fig.4(a)), a transformation from the amorphous structure into polycrystalline one was observed when the growing film thickness of about 100 nm was reached. It can be seen in Fig.4(a) that as the thickness of the film increases grains growing from initially sparsely located nuclei become wider and form a polycrystalline film with a columnar structure. The RHEED pattern of the *poly*–Si film grown at 450°C was similar to that presented in Fig.1(b) for the 50 nm thick Si film grown at 500°C. The Si surface has noticeable undulation that allows us to estimate the lateral size of the grains near the film surface as exceeding 100 nm. The *poly*–Si film contains defects that are highly likely grain boundaries, stacking faults, and twins (multiple twins), which appear as slanted or vertical parallel lines running through the grain [33–35]. Tilt boundaries that are formed by stacking faults or a Σ3 with [110] rotation axis twin boundary ($\theta = 70.53°$) in {111} planes with the lowest energy of formation [34,36] do not introduce any large bond distortions and are prevalent structural defects in contrast to twist boundaries [37].

In the sample grown at 500°C, the polycrystalline columnar structure penetrates the entire film (Fig.4(b)). This *poly*–Si layer is more defective than that grown at 450°C. At the *poly*–Si/$Si_3N_4$ interface, a large number of small grains (nuclei) simultaneously form. In the lower half of the film (Fig.4(b)), a gradual transition is observed from the structure with a large number of small grains to that composed of fewer grains of a larger size. In the near-surface area, the grain lateral sizes are about 100 nm and less. In the *poly*–Si films grown at 650°C, these sizes exceed 150 nm (the TEM image is not shown).

The columnar structure is also observed in the *poly*–Ge layers. TEM images of the samples grown at 400°C and 500°C are presented in Fig.5. In the both films, the most defective region is located near the *poly*–Ge/$Si_3N_4$ interface; some defects extend to the top of the layer, however. In the near-surface area, the grain lateral sizes grow with the film deposition temperature; they reach 100 nm at the growth temperature of 400°C (Fig.5(a)) and exceed 200 nm at 500°C (Fig.5(b)). As the growth temperature of the film increases, the surface of the layer becomes very rough and matte to the eye.

---

[1] Additional TEM images of similar films can be found in Ref. [32].

The results of RHEED, XRD and TEM show that the most defective areas with incoherent grain boundaries are located in the region near the interface with the substrate. The grains with the largest lateral sizes are situated in near-surface domains and these sizes increase with the film growth temperature. The extended defects such as grain boundaries, twins (multiple twins, twin boundaries) and stacking faults formed in the *poly*-films might be considered as one of the reasons for the appearance of regions determined by means of XRD as coherent scattering domains.

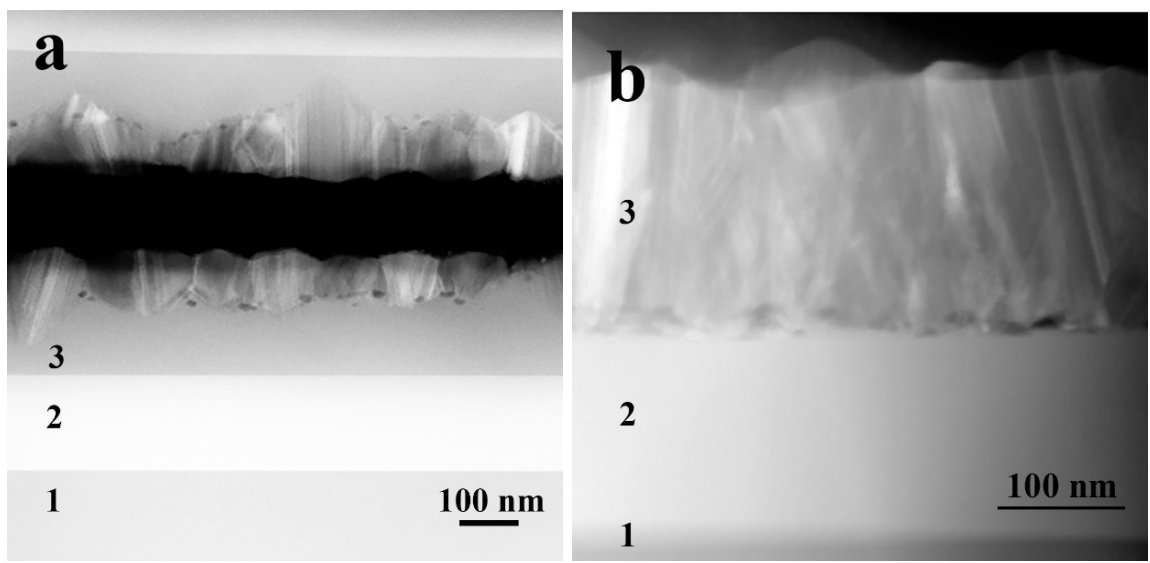

Fig.4. STEM images of the samples with the 200 nm thick Si films deposited at 450°C (a) and 500°C (b); the numbers denote (1) $SiO_2$, (2) $Si_3N_4$ and (3) the deposited Si film; the dark horizontal stripe in panel (a) is a layer of epoxy between two pieces of the same sample mounted symmetrically opposite each other on a holder.

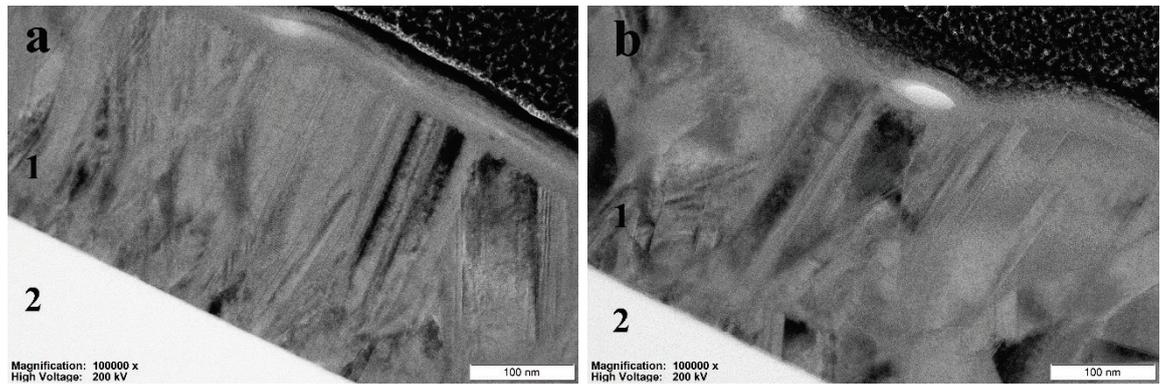

Fig.5 TEM images of samples with the 200 nm thick Ge films deposited at 400°C (a) and 500°C (b); the numbers denote (1) *poly*–Ge and (2) $Si_3N_4$ layers; a protective layer of Pt deposited at room temperature before the sample was thinned is visible on the surface.

## 3.2. Raman spectroscopy
### 3.2.1. Polycrystalline Si

Several samples with *poly*–Si films having different structural properties have been prepared for the study: (1) a 200 nm thick layer grown at 450°C (N1) (Fig.4(a)) composed of lower amorphous and upper polycrystalline parts; (2) entirely polycrystalline 200 nm thick (N2) and 50 nm thick (N3) films grown at 500°C (Fig.4(b)) and 200 nm thick one (N4) deposited at 650°C. The Raman spectra of all the samples contain a strong narrow vibrational band around 520 $cm^{-1}$ assigned to the TO phonon of crystalline silicon (TO(*c*-Si)) but only the sample N1 exhibits a broad peak around 480 $cm^{-1}$ associated with the TO phonons of $\alpha$–Si [38]. The Raman spectra obtained at the sample N1 using different excitation lasers are depicted in Fig.6. The spectrum obtained using a 405-nm laser is related to the upper polycrystalline part of the film. The spectrum obtained using a 785-nm laser contains an additional 2TA-phonon peak characteristic of crystalline silicon. However, in the spectra obtained using 532-nm and 633-nm lasers, this peak is not clearly expressed. Thus, one can conclude that laser radiation with a wavelength of 785 nm passes through the deposited Si film in the sample N1 and the TO(c-Si) band is enhanced due to the contribution of the substrate.

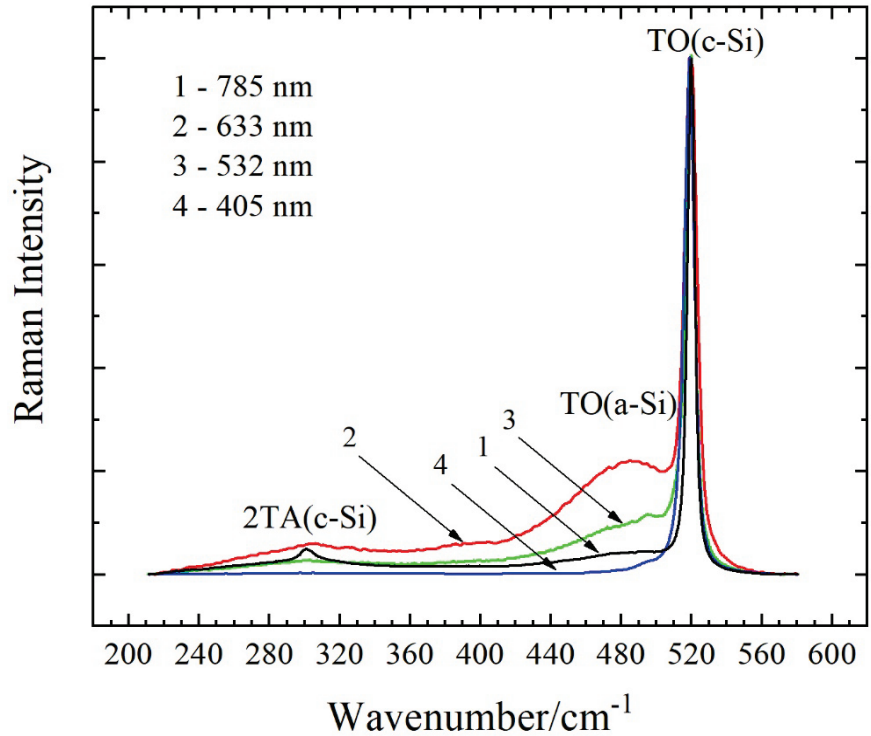

Fig.6 Comparison of the Raman spectra of the 200 nm thick Si layer grown at 450°C (N1) for different wavelength of the exciting laser; the spectra are normalized to the maximum of the strongest band peaked at 520 cm$^{-1}$.

Fig.7(a) demonstrates the Raman band at ~520 cm$^{-1}$ related to the polycrystalline part of the sample N1, recorded using a 405-nm laser, and a result of its analysis. This spectrum has been decomposed into three components: a strong vibrational band peaked at ~519 cm$^{-1}$ (Peak 1) and broad weaker bands peaked around 518 cm$^{-1}$ (Peak 2) and 500 cm$^{-1}$ (Peak 3) have been obtained. These peak components have been observed in the spectra of all the samples. Peak 1 best fits the Voigt function and its position is similar to that of the TO(*c*-Si) vibration band obtained for the Si$_3$N$_4$/SiO$_2$/Si (001) substrate. Peak 2 and Peak 3 are best approximated by Gaussian profiles. We were unable to deconvolute the band at ~520 cm$^{-1}$ into two components, i.e. combine Peak 1 and Peak 2 into one wide symmetrical peak. Therefore, for now we assume beforehand that these are two separate bands [39]. The analysis of Raman spectrum obtained using a 532-nm laser, the radiation of which penetrates the lower amorphous part of the N1 film, has revealed the same peaks as those shown in Fig.7(a) and one additional broad peak around 475 cm$^{-1}$ (Peak 4) corresponding to the TO($\alpha$–Si) vibrations (Fig.7(b)).

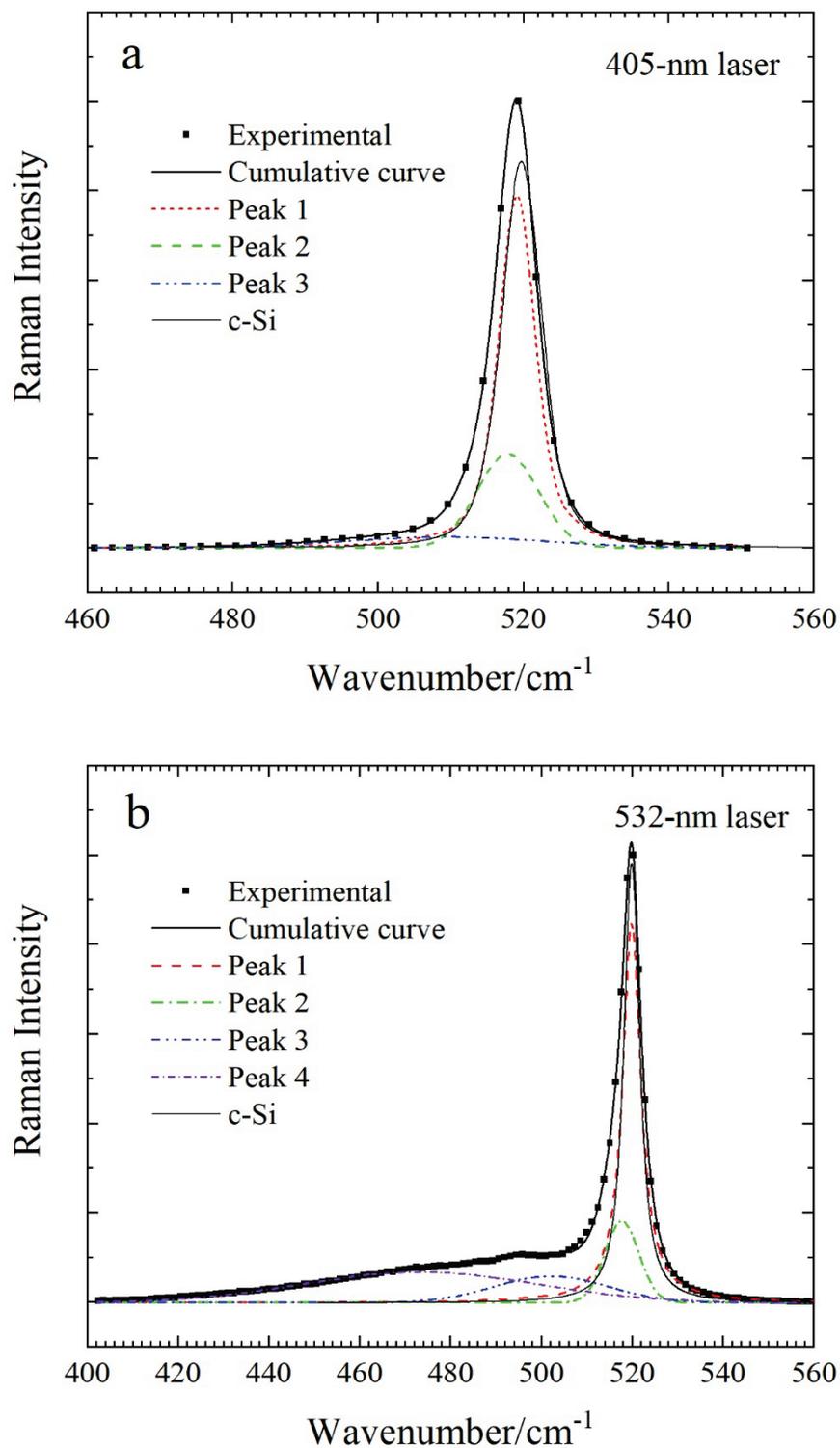

Fig.7 Deconvolution of Raman spectra of the 200 nm thick Si layer grown at 450°C (N1) in the vicinity of the TO(*c*-Si) vibration band peaked at 520 cm$^{-1}$; the *c*-Si spectrum relates to the Si$_3$N$_4$/SiO$_2$/Si(001) substrate; the laser wavelengths are 405

(a) and 532 nm (b); the spectra are normalized to the maximum of the strongest peak at 520 cm$^{-1}$.

In Fig.8 the changes in normalized peak areas of the components determined for the spectra obtained using different lasers are depicted. The normalized peak areas have been calculated as $S_{Peak\ N}/(S_{Peak\ 1}+S_{Peak\ 2}+S_{Peak\ 3})$, where $N = 1, 2$ or $3$. For all samples, a sharp increase in the area of Peak 1 was observed when using a laser with the wavelength of 785 nm, which is due to the strong influence of the substrate. In the Raman spectra of the samples N1, N2 and N3, Peak 1 displays similar parameters: its position practically coincides with that of the TO(*c*-Si); its full width at half maximum (FWHM) exceeds that in the reference sample by more than 1 cm$^{-1}$. With increasing influence of light scattering in the substrate, Peak 1 narrows and its FWHM coincides with that of the TO(c-Si) band in the reference sample. As for Peak 3, its specific area increases with approaching to the *poly*–Si/substrate interface. It can be assumed that the intensity of Peak 3 is related to the structure of the film: grain boundaries and the interface with amorphous Si (in the sample N1). The area of Peak 2 is always greater than that of Peak 3 in the domains situated near the surface of the 200 nm thick films (the samples N1 and N2 at 405-nm laser excitation in Fig.8a), which are less defective (Fig.4). As the laser radiation penetrates deeper into the film, the Peak 2 specific area decreases. For the samples N2 and N3 with different thicknesses (Fig.8b), the $S_{Peak\ 2}/S_{Peak\ 3}$ area ratio is less in the case of the 50-nm *poly*–Si film than in the case of the 200-nm one due to greater structural disorder in the sample N3, which affects the intensity of Peak 3. Completely different results are observed for the sample N4 (Fig.8a).

The Raman spectrum obtained using a 405-nm laser consists of two components: strong Peak 1 and almost vanishing Peak 3. The deconvolution of the Raman spectrum is presented in Fig.9. When using other lasers passing through the *poly*–Si film, Peak 2 again appears (Fig.8a). The Peak 1 FWMH exceeds that of the TO(*c*-Si) band in the *c*-Si spectrum by more than 2 cm$^{-1}$. With increasing scattering in the substrate, FWHM of Peak 1 becomes smaller; however, the peak remains wider than that in the c-Si spectrum. The intensity of Peak 3 increases with increasing laser wavelength. We assume that Peak 2 is absent because the near-surface domains consist of larger grains than those deep in the film.

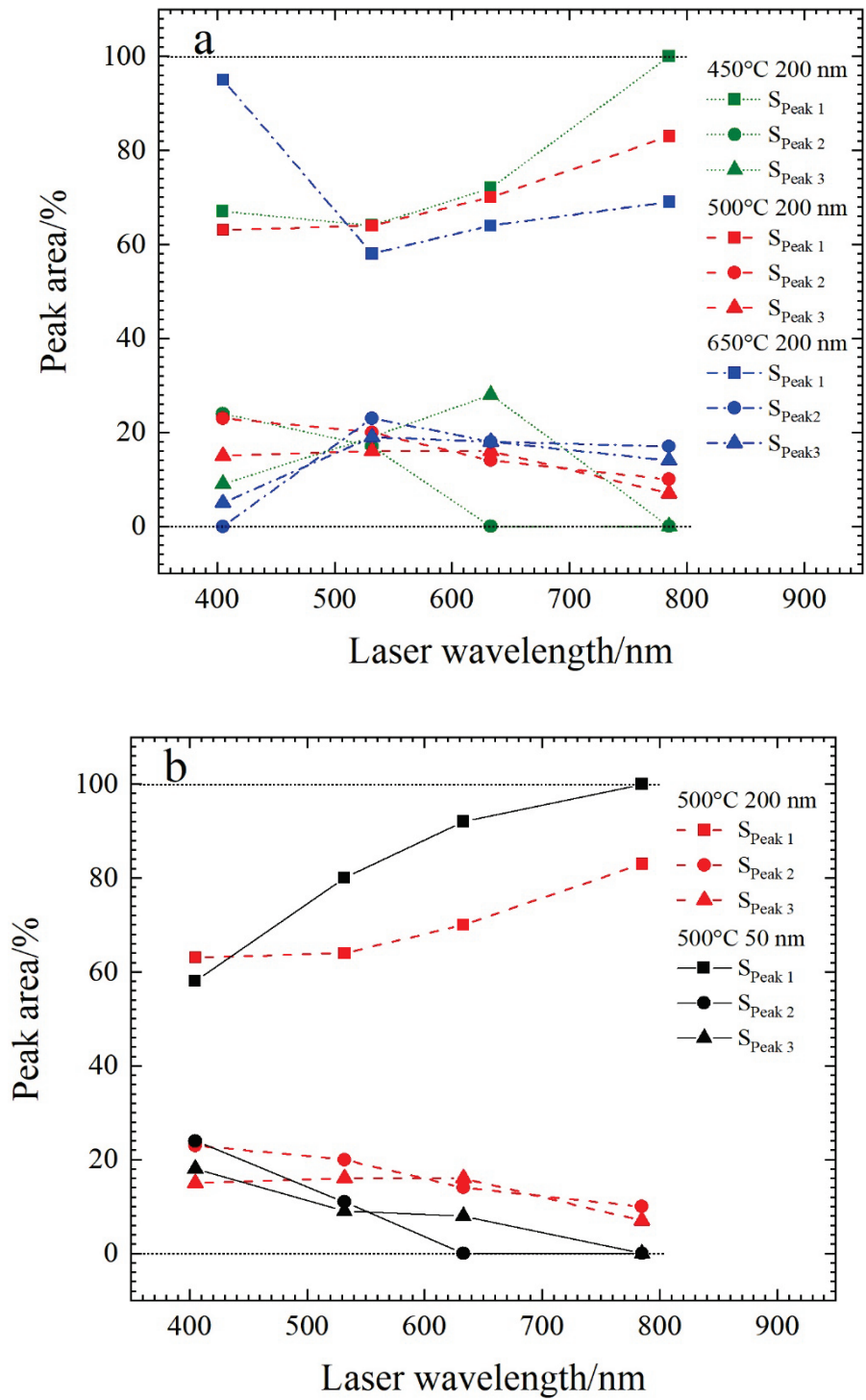

Fig.8 The change in the specific areas of peak components calculated for the spectra obtained using different lasers: (a) the samples with the 200 nm thick *poly*–Si films

grown at 450 (N1), 500 (N2) and 650°C (N4); (b) the samples with the 200 nm (N2) and 50 nm (N3) thick *poly*–Si films grown at 500°C; notations in the figure are as follows: 450°C, 500°C and 650°C are the growth temperatures; 200 nm and 50 nm are the thicknesses of the *poly*–Si films; square symbols correspond to Peak 1, circles to Peak 2 and triangular to Peak 3; a value equal to zero means that this peak component is absent; the lines are given as a guide to the eye.

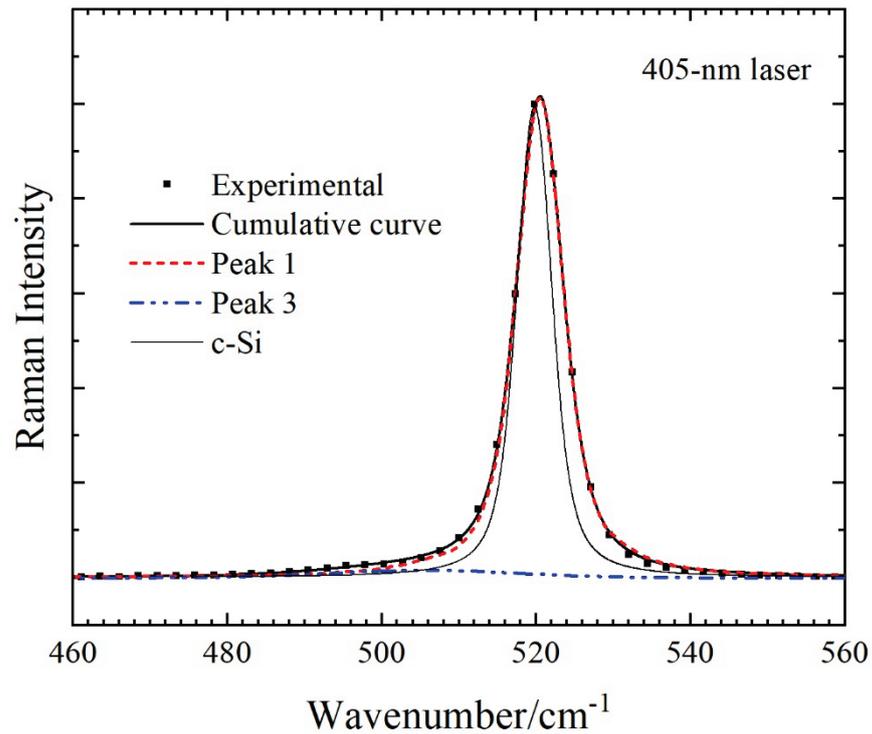

Fig.9 Deconvolution of the Raman spectrum of the 200 nm thick Si layer grown at 650°C (N4) in the vicinity of the TO(*c*-Si) vibration band peaked at 520 cm$^{-1}$; the *c*-Si spectrum is related to the $Si_3N_4/SiO_2/Si$ (001) substrate; the laser wavelength is 405 nm; the spectrum is normalized to the maximum of the strongest component peaked at 520 cm$^{-1}$.

Thus, the Raman spectra of the *poly*–Si films have the vibrational band around 520 cm$^{-1}$ consisting of three components.

The main conclusions are as follows. Firstly, the broad peak about 500 cm$^{-1}$ is attributed to severe structural disorder originating from GBs in the *poly*–Si film or at the α-Si/*poly*–Si interface. This Raman peak has been presented in experimental studies [1,18,40–44] as well as computational investigations of twist GBs and an *α*–Si/*c*-Si interface [45,46]. The peak area increases as the *poly*–Si/substrate interface is approached. According to Ref. [47], the distance, at which short-range order in the disturbed area of grain boarders is kept, might be estimated as ~1.2 nm. Secondly, there is a relationship between two neighboring peaks at 519 cm$^{-1}$ (Peak 1) and 518 cm$^{-1}$ (Peak 2). Peak 2 disappears if Peak 1 is strongly broadened. Thirdly, the intensities of Peak 2 and Peak 3 show similar behavior when changing the laser wavelength. We suppose that the bands at about 519 cm$^{-1}$ (Peak 1) and 518 cm$^{-1}$ (Peak 2) could be treated as one asymmetrical peak in the phonon confinement model bearing in mind that GBs and extended defects might be considered as the reason for phonon confinement.

### 3.2.2. Polycrystalline Ge

Two sets of the samples were prepared. The first set comprised *poly*–Ge films with the thickness of 50 nm, 100 nm and 200 nm grown at 500°C on a Si$_3$N$_4$/SiO$_2$/Si (001) substrate. The second one included samples with 200 nm thick *poly*–Ge film deposited at the temperature ranging from 400 to 600°C. All these films were entirely polycrystalline. Raman spectra have been obtained using 532, 633 and 785-nm lasers.

In the Raman spectra, regardless of the thickness of the sample grown at 500°C, the strongest vibrational band is peaked at 300 cm$^{-1}$; it is assigned to a TO phonon band of crystalline germanium (TO(c-Ge)). The result of deconvolution of the vibrational band excited using a 633-nm laser for the sample with a 200 nm thick *poly*–Ge film is shown in Fig.10. It is composed of a strong narrow symmetrical peak at ~300 cm$^{-1}$ (Peak 1) and a broad weak band near 290 cm$^{-1}$ (Peak 2). Peak 1 and Peak 2 are fitted by the Voigt and Gaussian functions, respectively. Within the measurement error, the position of Peak 1 coincides with that of the TO(*c*-Ge) in the reference sample [48,49]. FWHM of Peak 1 exceeds that of the TO(*c*-Ge) band by more than ~1–1.5 cm$^{-1}$. The change in the specific area of Peak 2 depending on the laser wavelength is shown in Fig.11. The radiation from all lasers passes through the 50 nm thick *poly*–Ge film. (The penetration depth of the exciting laser radiation is estimated by the appearance of the TO(*c*-Si) band in the spectrum.) The radiation of both 633-nm and 785-nm lasers passes through the 100-nm film. Regarding 200 nm thick sample, only the radiation of a 785-nm laser penetrates the substrate. Owing to a small penetration depth of 532-nm laser light, the contribution of the surface to the Raman spectra obtained when excited by this radiation is significant. It is seen in Fig.11 that the specific area of Peak 2 increases in domains near the interface with the substrate. We assume that the vibrational band at ~290 cm$^{-1}$ (Peak 2) may be associated with incoherent grain boundaries in *poly*–Ge [12,50] and have the same nature as the band at ~500 cm$^{-1}$ in the Raman spectra of *poly*–Si (see Section 3.2.1).

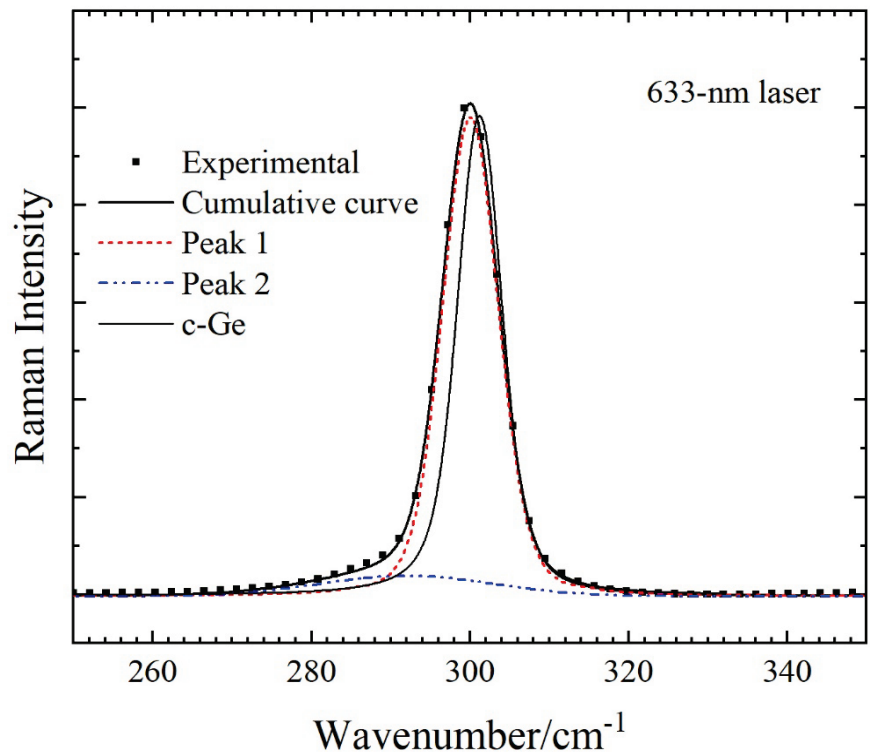

Fig.10 Result of the deconvolution of Raman spectra around the peak at 300 cm$^{-1}$ of the sample with the 200 nm thick *poly*–Ge film grown at 500°C; the spectrum of *c*-

Ge is related to crystalline germanium; the spectra are normalized to the maximum of the strongest peak at 300 cm$^{-1}$.

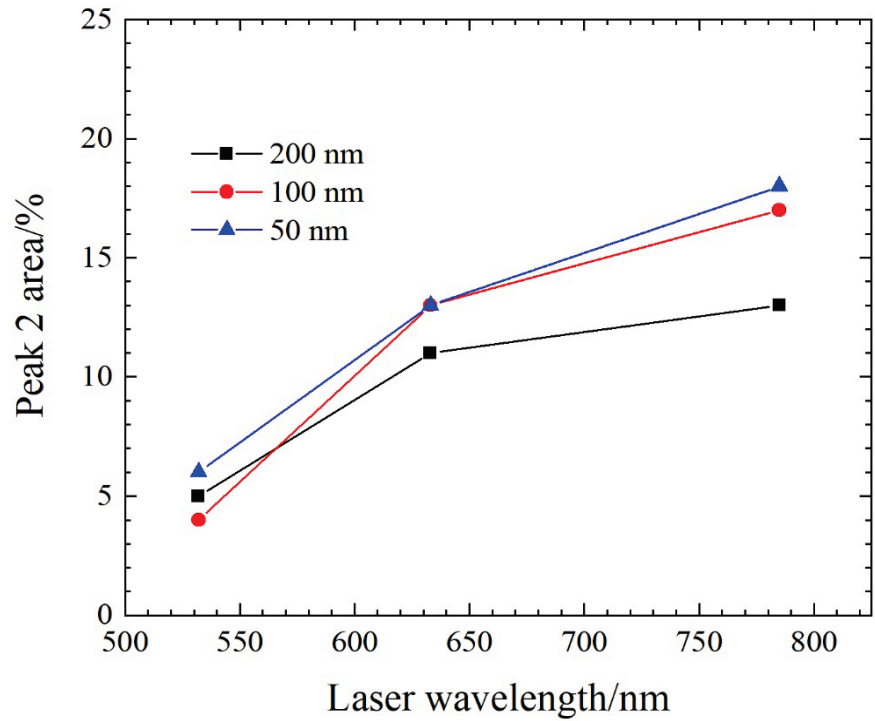

Fig.11 Specific area of Peak 2 (~290 cm$^{-1}$) in the spectra vs an exciting laser radiation wavelength; notations on the figure are as follows: squares correspond to the 200 nm thick *poly*–Ge film, circles to the 100 nm thick one and triangles to the 50 nm thick one; the lines are given as a guide to the eye.

The 200 nm thick *poly*–Ge films grown in the temperature range from 400 to 600°C have different polycrystalline structure that follows from the results of the TEM study (Fig.5). Raman spectra obtained using the 785-nm laser radiation, the only one passing through all *poly*–Ge layers, are presented in Fig.12. An increase in the intensity ratio $I_{TO(Ge)}/I_{TO(Si)}$ is observed with raising film deposition temperature (see the insert in Fig.12) that is connected with an improvement of *poly*–Ge structure i.e., with increasing order throughout the depth of the film, therefore, with a decrease in Rayleigh scattering of the exciting light and its deeper penetration into the film. The change in the specific area of Peak 2 depending on the film growth temperature is presented in Fig.13. The specific area of Peak 2 decreases in near-surface region of the films, where column grains have a larger lateral size. The higher the film growth temperature, the smaller contribution of Peak 2 into the overall peak. These results support the above assumption that the vibration band peaked at ~290 cm$^{-1}$ is associated with defects, the density of which is the largest near the interface between the *poly*–Ge film and the substrate.

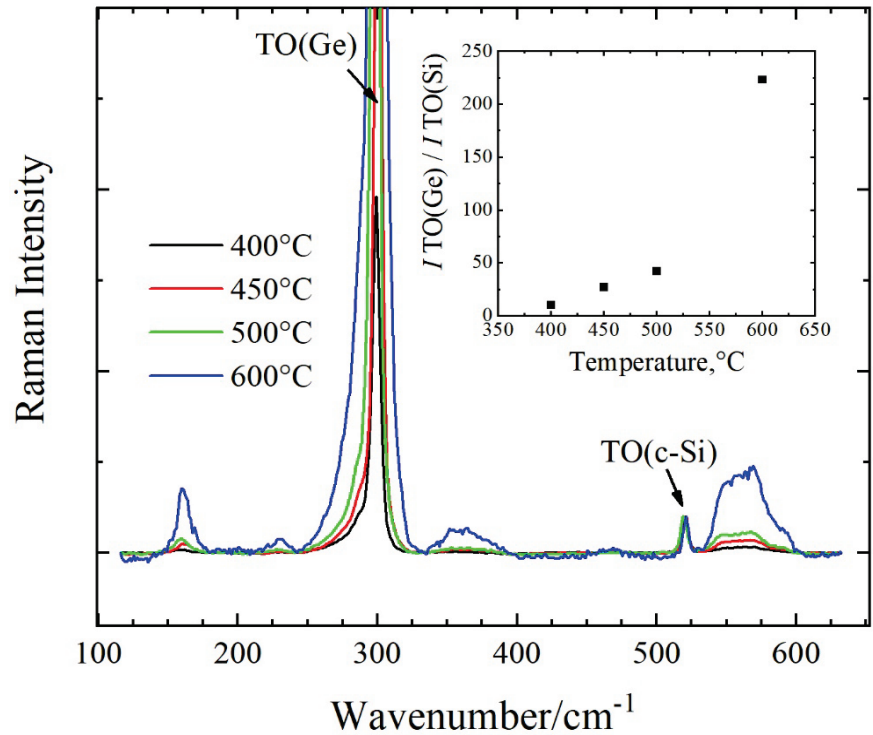

Fig.12 Raman spectra of the 200 nm thick *poly*–Ge films grown at the temperatures of 300°, 450°, 500° and 600°C; all the spectra have been acquired using a 785-nm laser and normalized to the TO(*c*-Si) peak intensity; the insert shows an intensity ratio $I_{TO(Ge)}/I_{TO(Si)}$ as a function of the film deposition temperature.

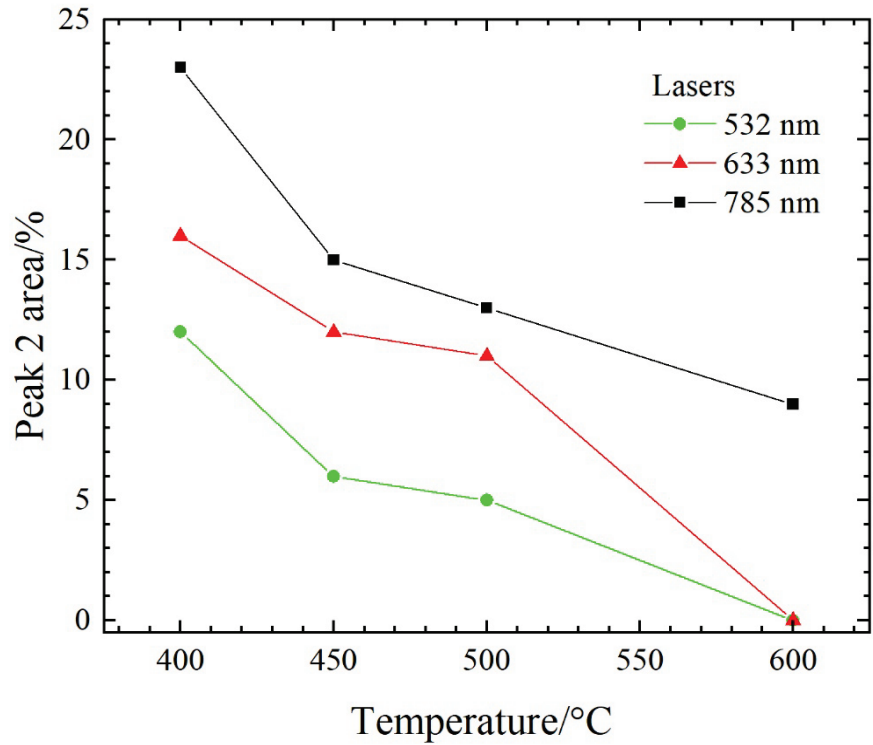

Fig.13 Dependence of the specific area of Peak 2 on the film growth temperature; 200 nm thick *poly*–Ge films deposited at 400, 450, 500 and 600°C; a specific area value of zero means that Peak 2 is absent from the band; the lines are given as a guide to the eye.

**3.2.3. Simulation of Raman intensity**

Raman intensity has been simulated in the framework of a spatial correlation model proposed by Richter [1] and modified by Campbell and Fauchet [2]. To model the Raman spectrum, it is necessary to determine the shape of the nanocrystals that make up the film. The structure of the obtained polycrystalline films cannot be considered as columnar since the condition of a strong difference between the diameter and length of the column is not met. For the sake of certainty, we adopt the spherical shape of the nanocrystals forming *poly*–Si and *poly*–Ge films. We assume that elastic stresses do not contribute to the position of peaks in the spectra. We also neglect the possible size distribution of crystallites, since it could not be determined.

The Raman intensity is defined as [1,2]

$$I(\omega) = \int \frac{|C(0,q)|^2}{[\omega - \omega(q)]^2 + \left(\frac{G_0}{2}\right)^2} d^3q, \tag{1}$$

where $C(0,q)$ is a Fourier coefficient of the confinement function, $\omega(q)$ is the phonon dispersion, $q$ is the phonon momentum, $G_0$ is the linewidth of the TO phonon band in the reference c-Si or c-Ge samples and reflects the phonon lifetime. The $d^3q$ is treated as $\sim q^2 dq$ for spherical nanocrystals [5,10]. In the c-Si and c-Ge reference samples, the Lorentzian curve shape of the TO(c-Si) or TO(c-Ge) bands have been observed only when using the 532 nm laser. In cases of the 405, 633 and 785-nm excitation sources, the curve shape has been fitted by the Pseudo-Voigt function (the linear combination of the Lorentzian and Gaussian functions) or the Voigt one. In Fig.S1(a) (Supporting Information), the Raman spectrum of the c-Si reference sample obtained using a 532-nm laser and fitted by the Lorentzian function is depicted. In Fig.S1(b) (Supporting Information), the Raman spectrum has been measured using a 405-nm laser and treated using different functions. The best results are observed using the Pseudo-Voigt and Voigt functions having an additional Gaussian component. The Voigt function best fits the spectra of samples, in which Peak 2 is absent in the TO band, such as those shown in Fig.9 and Fig.10. We believe that the expression (1) should be modified by replacing the Lorentzian function with the Pseudo-Voigt one in order to take into account the structural disorder in the polycrystalline films, leading to the broadening of the peaks, as well as a possible instrumental effect.

$$I(\omega) = \int_0^1 |C(0,q)|^2 \left( \frac{2k\, G_L}{\pi(4[w - w(q)]^2 + G_L^2)} \right.$$
$$\left. + (1-k)\frac{\sqrt{4ln2}}{\sqrt{\pi}\, G_G} \exp\left\{ -\frac{4ln2}{G_G^2}[w - w(q)]^2 \right\} \right) d^3q$$
$$\tag{2}$$

where $G_L$ and $G_G$ are the linewidths of the Lorentzian and Gaussian components, respectively, and k is a relative fraction. In the calculation, the value of $G_L$ was taken as constant one and equal to the $G_L$ value in the spectrum of the c–Si reference.

The Si and Ge optical phonon dispersion curves are accepted as isotropic, LO and TO branches are taken as matching [7,18] and the dispersion relation of the TO phonon in the [100] direction is [51,52]

$$\omega(q) = \sqrt{A + B\cos\left(\frac{\pi q}{2}\right)} + Z, \quad (3)$$

where $Z$ is an adjustment parameter, arising from comparison with experimental curves, and $q$ is in the range [0;1]. For the Si films, $A = 1.714 \times 10^5$ and $B = 1.000 \times 10^5$ are used, and for the Ge films $A = 0.69 \times 10^5$ and $B = 0.195 \times 10^5$ [14,52].

The Fourier coefficient is [5,14,25]

$$|C(0,q)|^2 = \exp\left\{-\frac{1}{2}\left(\frac{qL}{\alpha a}\right)^2\right\}, \quad (4)$$

where $L$ is an average crystal domain size in polycrystalline film, $a$ is the lattice constant ($a_{Si} = 0.54307$ nm and $a_{Ge} = 0.566$ nm). The choice of the value of the parameter $\alpha$ is the main issue for spectrum modeling. In the classical works of Richter [1] and Campbell and Fauchet [2], it was $\sqrt{2}$ and $\pi\sqrt{8}$, respectively, that was determined by the confinement envelope function $W(r, L) = \exp\{-(\alpha r/L)^2\}$ ($r$ is the radial coordinate), which was Gaussian. The using of other values of the parameter $\alpha$ was discussed in the works [1,2,5,11,14,19–25]. In all the cases, the parameter $\alpha$ was selected individually. Fig.S2 (Supporting Information) shows, how different values of the parameter $\alpha$ influence a peak shift and a line broadening. The calculation is carried out according to the equation (2) at $L = 30$ nm and $k = 1$ ($\alpha = 4.2$, found in the current work). As a result, the use of the parameter $\alpha = \pi\sqrt{8}$ [2] in the Si structures leads to such a peak shift and broadening of the spectrum that are not observed in our *poly*-Si films, and the spectrum simulated with $\alpha = \sqrt{2}$ [1] coincides with that of *c*-Si (Fig.S2(a) (Supporting Information)). In the Ge nanostructures, there are less noticeable changes in these parameters than in the Si ones (Fig.S2(b) (Supporting Information)). Since the model is phenomenological, the difference between Si and Ge might be observed due to $a$ (lattice constant) and $w(q)$ parameters. We have calculated the parameter $\alpha$ using XRD data on the size of coherent scattering domains $D = 26$ nm in *poly*–Si, the Raman spectrum of which is given in Fig.S3 (Supporting Information). This sample is the 50 nm thick *poly*–Si film grown at 500°C; its Raman spectrum has been obtained using a 405-nm laser, the radiation of which passes through the entire *poly*–Si film, but the contribution of the substrate is negligible. The asymmetric peak taken for fitting was obtained by summing Peak 1 and Peak 2. We have calculated that the parameter $\alpha = 4.2$. Fig.S4 (Supporting Information) presents the result of this simulation. For comparison, the peak of *c*-Si is depicted in the position corresponding to one of the curves simulated with $L = 200$ nm when any influence of the phonon confinement already disappears. In the *poly*–Si film, the peak position shifts down by ~1.9 cm$^{-1}$ and FWHM increases by ~2.4 cm$^{-1}$ relative to the TO(*c*-Si) peak (Fig.S4 (Supporting Information)). The Raman spectrum fitted using the simulated peak is presented in Fig.14. In the same way, we have treated the Raman spectra of the 200 nm thick *poly*–Si films grown at 500°C and 650°C, obtained using a 633-nm laser, the radiation of which passes through almost the entire *poly*–Si film. In the first case (500°C), we obtained $\alpha = 4.2$ at $L = 27$ nm, while in the second case (650°C), $\alpha = 4.2$ at $L = 30$ nm that corresponds to the XRD results in both cases. As a result, we have adopted the parameter $\alpha \sim 4.2$ for further modeling. The values of $L$ and $\alpha$ yielded by the simulation of the Raman spectrum and the value of $D$ obtained from the XRD study are gathered in Table 1. The values of $L$ and $D$ become equal when the influence of the domains located near the *poly*–Si/substrate interface on Raman spectra is significant. In the near-surface regions, $L$ values obtained by modeling for the 200 nm thick films (at a 405-nm exciting laser) become larger, which correlates with the TEM results.

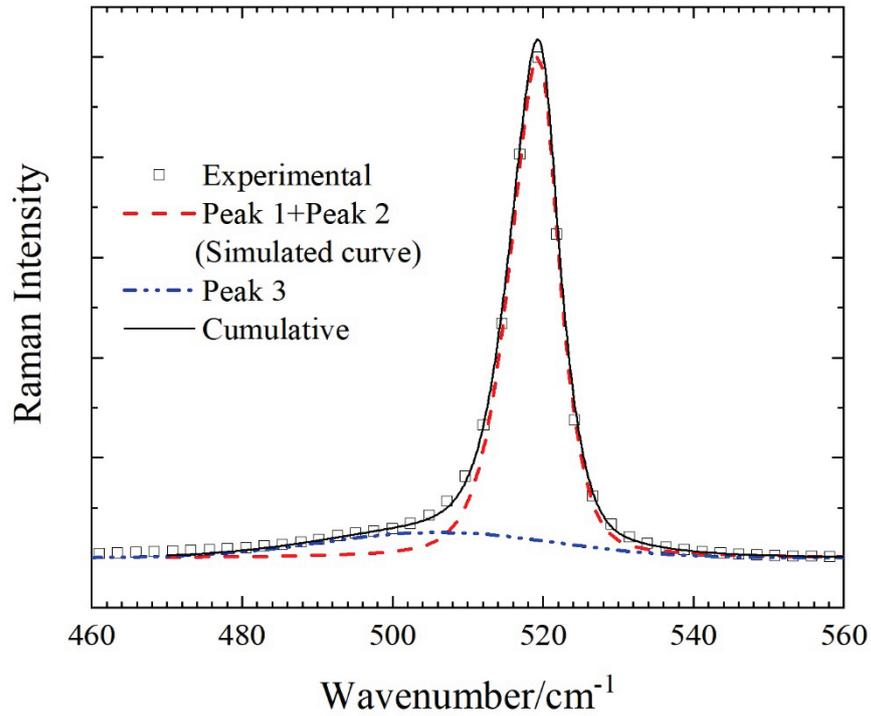

Fig.14 Deconvolution of the Raman spectrum with one asymmetrical peak (simulated curve in Fig.S4 (Supporting Information)) and one symmetrical peak (Peak 3 in Fig.S3 (Supporting Information)) of the 50 nm thick *poly*–Si film grown at 500°C (N3), a 405-nm laser; the spectrum is normalized to the maximum of the strongest peak at 520 cm$^{-1}$.

Table 1 Comparison of the parameters $\alpha$ and $L$ obtained by modeling Raman spectra with the dimensions of coherent scattering domains ($D$) obtained by XRD.

| Sample | Film growth temperature, °C | Film thickness, nm | Laser wavelength, nm | Raman | | XRD |
|---|---|---|---|---|---|---|
| | | | | $\alpha$ | $L$, nm | $D$, nm |
| N1 | 450 | 200 | 405 | 4.2 | 36 | |
| N1 | 450 | 200 | 532 | 4.2 | 36 | |
| N3 | 500 | 50 | 405 | 4.2 | 26 | 26 |
| N2 | 500 | 200 | 405 | 4.2 | 27 | |
| N2 | 500 | 200 | 532 | 4.2 | 28 | 26 |
| N2 | 500 | 200 | 633 | 4.2 | 27 | |
| N4 | 650 | 200 | 405 | 4.2 | >70 | |
| N4 | 650 | 200 | 532 | 4.2 | 28 | 30 |
| N4 | 650 | 200 | 633 | 4.2 | 30 | |

The Raman spectra of the *poly*–Ge films do not have any features connected with phonon confinement. Peak 1 is a broad symmetrical band and its position practically is not shifted relative to that of the TO(*c*-Ge) band in the reference *c*-Ge (Fig.10). We have found that the maximal values of the parameter $\alpha$ are in the range from 2 to 3 when Raman peaks are symmetrical for the value of $L = D$ and the laser radiation passes through the *poly*–Ge film. In [53], in the Raman spectra of Ge nanowires with a core coated with an oxide shell with a diameter of 20 to 51 nm, an asymmetrical broadening without a downward shift was observed. Thus, we believe that the larger nanocrystal size is the reason for the lack of phonon confinement in the *poly*–Ge films, and a symmetrical broadening of the peak may be due to structural disorder reducing the phonon lifetime.

Thus, in the model considered in Refs. [1,2], a low value of the parameter $\alpha$ suggests that the phonon is not confined within the nanocrystal and can propagate beyond the grain boundary, since the intragranular and intergranular regions consist of the same substance. The different structures of these domains (symmetry breaking, defects, etc.) affect the phonon wavelength and the Raman peak width.

To evaluate the possibility of phonon propagation across twin grain boundaries, we have simulated phonon dispersion and density of phonon states (DOS) for structures with multiple Σ3 [110] twins. The simulation was performed using the CASTEP Density functional modeling (DFT) code [54] with GGA–PBE (Perdew–Burke–Ernzerhof) exchange correlation functional [55] and a density-functional perturbation theory and linear response approaches [56]. The parameters are described in more detail in Supporting Information. Unit cells were cut from crystalline *c*-Si (*c*-Ge) with the diamond structure. Si reference sample was a c-Si unit cell cut by the (111) cleavage plane. The multiple twin structure was composed of a sequence of Σ3 [110] twins (with the (111) boundary plane) that was made up of alternating c-Si unit cells of the (111) and ($\overline{111}$) cleavage planes. They are depicted in Fig.S5 (Supporting Information).

In Fig.S6 (Supporting Information), the first Brillouin zone of the supercell is shown. Fig.S7 and Fig.S8 (Supporting Information) demonstrate the results of modeling of phonon dispersion and phonon DOS. The densities of phonon states of these structures are depicted in Fig.15. It is seen that there is some discrepancy in the frequency range of optical phonons. In the multiple twin Si structure, the redistribution of density of phonon states is observed as the increase in DOS for phonons with lower wavenumbers around 400 cm$^{-1}$ relating to the LO peak and the decrease in DOS of high-frequency phonons near 475 cm$^{-1}$ relating to the TO peak (Fig.15(a)).

In Fig.15(b), the results of modeling for Ge structures are depicted. In this case, curves of density of phonon states practically overlap. Therefore, one can suppose that the barrier to phonon propagation between the multiple twin structure and *c*-Si (*c*-Ge) should be insignificant. We have compared the results of our simulation with those presented in Refs. [45,57–61]. The DFT modeling of the Si and Ge structures containing twinning (one (111) twin separated by three atomic bilayers) is presented in [57]. The phonon DOS distributions of c-Si (c-Ge) and twinning structures did not differ in that article. The results of the molecular dynamics simulations (MD) of the structures with some twist and tilt grain boundaries and α-Si/c-Si interfaces can be found in Refs. [45,58,59]. The simulated vibrational densities of states (VDOS) corresponding to the inner crystalline structure of the grain and the domains of grain boundaries differ significantly. For the intragrain regions, VDOS is related to c-Si structure, however, in GB domains, VDOS has intermediate view between c-Si and α-Si. Phonons are localized inside the grains when intragrain region is α-Si or composed of twin GDs with high misorientation angles. It was shown in [60] that twin boundaries could scatter high-frequency phonons.

The results of investigation of the phonon scattering carried out using a scanning transmission electron microscope (STEM) with electron energy loss spectroscopy (EELS) of Si containing symmetric Σ3 (111) and Σ9 (221) and asymmetric (111)|(115) grain boundaries are presented in Ref. [61]. The results of measurement of phonon DOS and MD DOS modeling for different regions of the polycrystalline Si were demonstrated in that article. It was shown that with an increase in the degree of disorder in the grain boundary region, decrease in the intensity of the optical phonon mode is observed. The authors of Ref. [61] associate these changes with a significant distortion of the bond angles. The Σ3 (111) grain boundary does not modify the phonon DOS, hence, our results do not contradict those ones. The Σ3 [110] does not introduce significant changes in the phonon DOS, therefore it has no significant influence on the Raman spectrum. Incoherent grain boundaries (intergrain area) having strong disordering in the Si (Ge) atom sites by analogy with α-Si or twin GDs with a high misorientation angles might result in redistribution of the phonon DOS that prevents their spreading beyond the grain. In Raman spectra, the appearance of the lower frequency peaks is observed as well as an asymmetrical broadening and a downward shift of the TO phonon band. If the grains are large enough and do not limit the phonon mean free path (MFP) then phonons are scattered by structural defects and the Raman spectrum broadens. This is what is observed in the *poly*–Ge films.

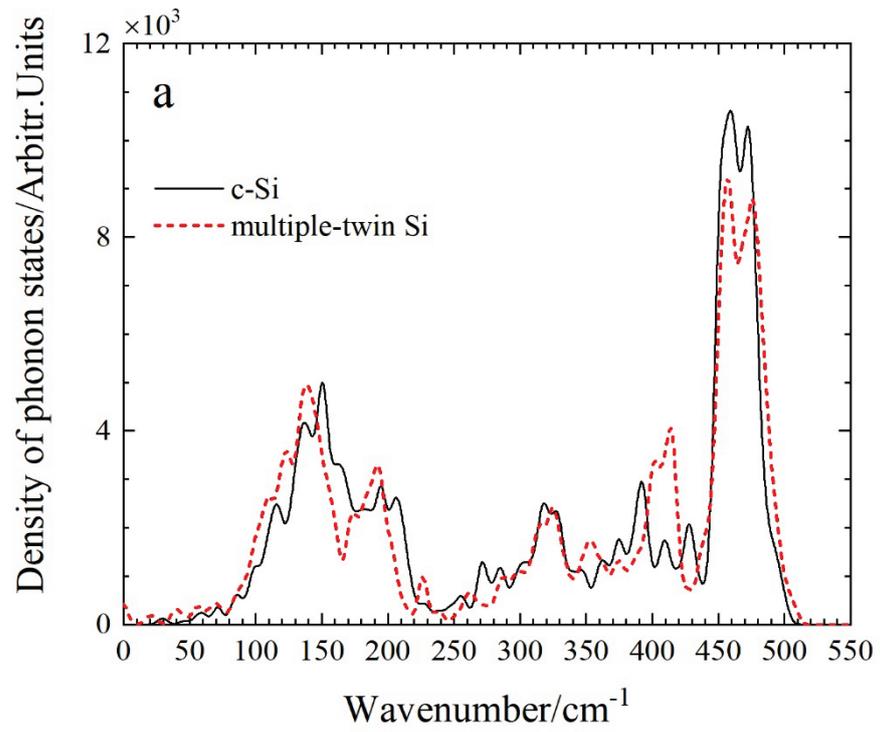

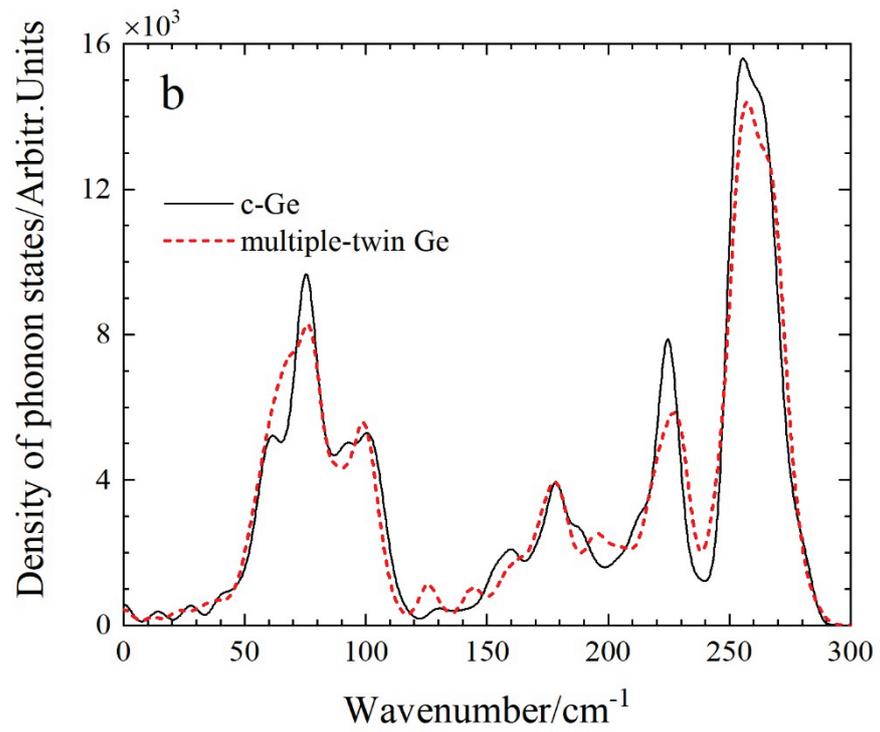

Fig.15 Simulated densities of phonon states for (a) *c*-Si and a multiple-twin Si structure and (b) for *c*-Ge and a multiple-twin Ge structure.

### 3.3. Conclusion

The results of RHEED, TEM, XRD and Raman spectroscopy studies of the polycrystalline Si and Ge films deposited on a dielectric $Si_3N_4/SiO_2/Si$ (001) substrates are presented in this work. It has been found that polycrystalline films have a columnar structure and consist of grains, the lateral size of which increases with increasing film thickness. *Poly*–Si films have been shown to be formed by grains of smaller size than *poly*–Ge ones. Since no preferential grain orientation was observed by XRD, this means that the film structure has a large number of incoherent grain boundaries, especially in the regions close to the interface with the substrate. In addition, in the intragranular region, there are other types of extended defects such as twins (multiple twins, twin boundaries). The Raman spectra of the films exhibit a number of features associated with their polycrystalline structure. The main vibrational bands of TO(*c*-Si) and TO(*c*-Ge) phonons around 520 and 300 cm$^{-1}$ have additional broad weaker peaks around 500 and 290 cm$^{-1}$, respectively, and changes in their intensities are due to severe structural disorder originating from GBs in the polycrystalline films and at the interface between the polycrystalline film and the substrate. Here, the scattering of radiation is similar to that in an amorphous material.

In the Raman spectra of the poly–Si films, an asymmetrical broadening and a downward shift of the TO(c-Si) phonon band have been observed. This peak could be fitted by two components. The first one is a strong vibrational band peaked at ~519 cm$^{-1}$ and the second one is a broad weaker band around 518 cm$^{-1}$. In the Raman spectra of the *poly*–Ge films, the TO(c-Ge) vibrational band shows only a symmetrical broadening and is fitted by a single peak. We believe that the downward shift and the asymmetrical broadening of the TO-phonon band at 520 cm$^{-1}$ in *poly*–Si may be associated with the phonon confinement. The source of the limitations is incoherent grain boundaries. We have modeled the Raman spectrum within the framework of the phonon confinement theory, taking into account the sizes of the coherent scattering domains estimated by the XRD method. The value of the parameter $\alpha$, which determines the length scale for the phonon confinement in the confinement envelope function $W(q) = \exp\{-(\alpha r/L)^2\}$, was determined to be 4.2, which corresponds to soft confinement of phonons at the nanocrystal boundary. We have simulated the DFT phonon dispersion and density of phonon states for a Si crystal consisting of multiple twins. The simulation results allow us to suggest that the barrier for phonon propagation between the multiple twin structure and c-Si should be insignificant. We suggest that symmetrical tilt twins could not be considered as the boundaries of the nanocrystal, in which the phonon confinement occurs.

The phonon confinement is absent in the *poly*–Ge films that might be due to both the larger grain size and the lower frequency of the TO(*c*-Ge) phonon compared to the TO(*c*-Si) one, for which grain boundaries are not barriers. The symmetrical broadening is associated with the scattering of phonons by defects and disordered structure. A similar symmetrical broadening was observed in *poly*–Si films when using lasers penetrating into the near-surface region consisting of grains with large lateral sizes.

**Conflict of interest statement**

The authors declare that they have no known competing financial interests or personal relationships that could have appeared to influence the work reported in this paper.

# Effect of extended defects on phonon confinement in polycrystalline Si and Ge films

Supplementary Materials

Larisa V. Arapkina[1], Kirill V. Chizh[1], Oleg V. Uvarov[1],
Valery V. Voronov[1], Vladimir P. Dubkov[1], Mikhail S. Storozhevykh[1],
Maksim V. Poliakov[2], Lidiya S. Volkova[2], Polina A. Edelbekova[2],
Alexey A. Klimenko[2], Alexander A. Dudin[2] and Vladimir A. Yuryev[1]

[1] Prokhorov General Physics Institute of the Russian Academy of Sciences,

38 Vavilov Street, 119991 Moscow, Russia

[2] Institute of Nanotechnology of Microelectronics of the Russian Academy of Sciences,

32A Leninsky Prospekt, 119991 Moscow, Russia

To study the effect of the laser wavelength on the shape of Raman scattering lines, the spectra of $c$–Si and $c$–Ge reference samples were recorded at various wavelengths of exciting laser radiation. Fig. S1 shows the experimental TO(c-Si) Raman band curves for 532 and 405-nm lasers. The Raman band obtained using a laser with a wavelength of 532 nm is best fitted by the Lorentzian profile with the coefficient of determination $R^2 = 0.99999$.

Another spectrum obtained using a 405-nm laser is fitted by: (1) the Lorentzian curve with $R^2 = 0.99254$, (2) the pseudo Voigt one (a liner combination of the Lorentzian and Gaussian functions) with $R^2 = 0.99986$ and (3) the Voigt one with $R^2 = 0.99996$. The best fitting parameters are reached using the pseudo Voigt function and the Voigt function, which are almost identical, however. We attribute this result with a possible instrumental influence on the TO peak shape and consider that it can be taken into account by using the pseudo Voigt function instead of the Lorentzian one in the expression for the Raman intensity $I(\omega)$. This replacement obviously simplifies further computations compared to using the Voigt function.

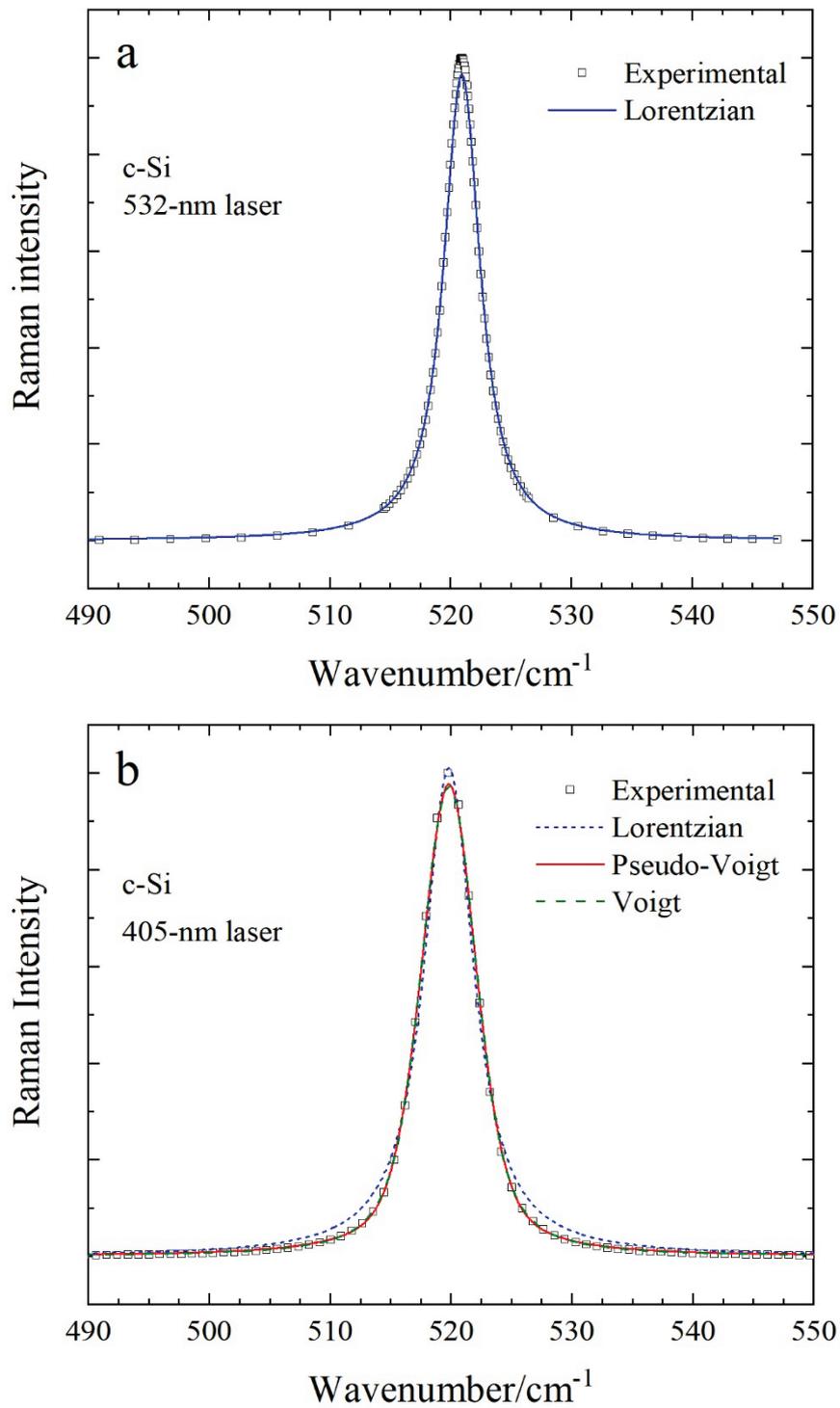

Fig. S1 Experimental and simulated curves of the TO(c-Si) Raman band obtained using 532 (a) and 405-nm (b) lasers; in panel (a), experimental dots are fitted by the Lorentzian curve, while in panel (b), they are fitted by the Lorentzian function, the pseudo Voigt function (a linear combination of the Lorentzian and Gaussian curves) and the Voigt function.

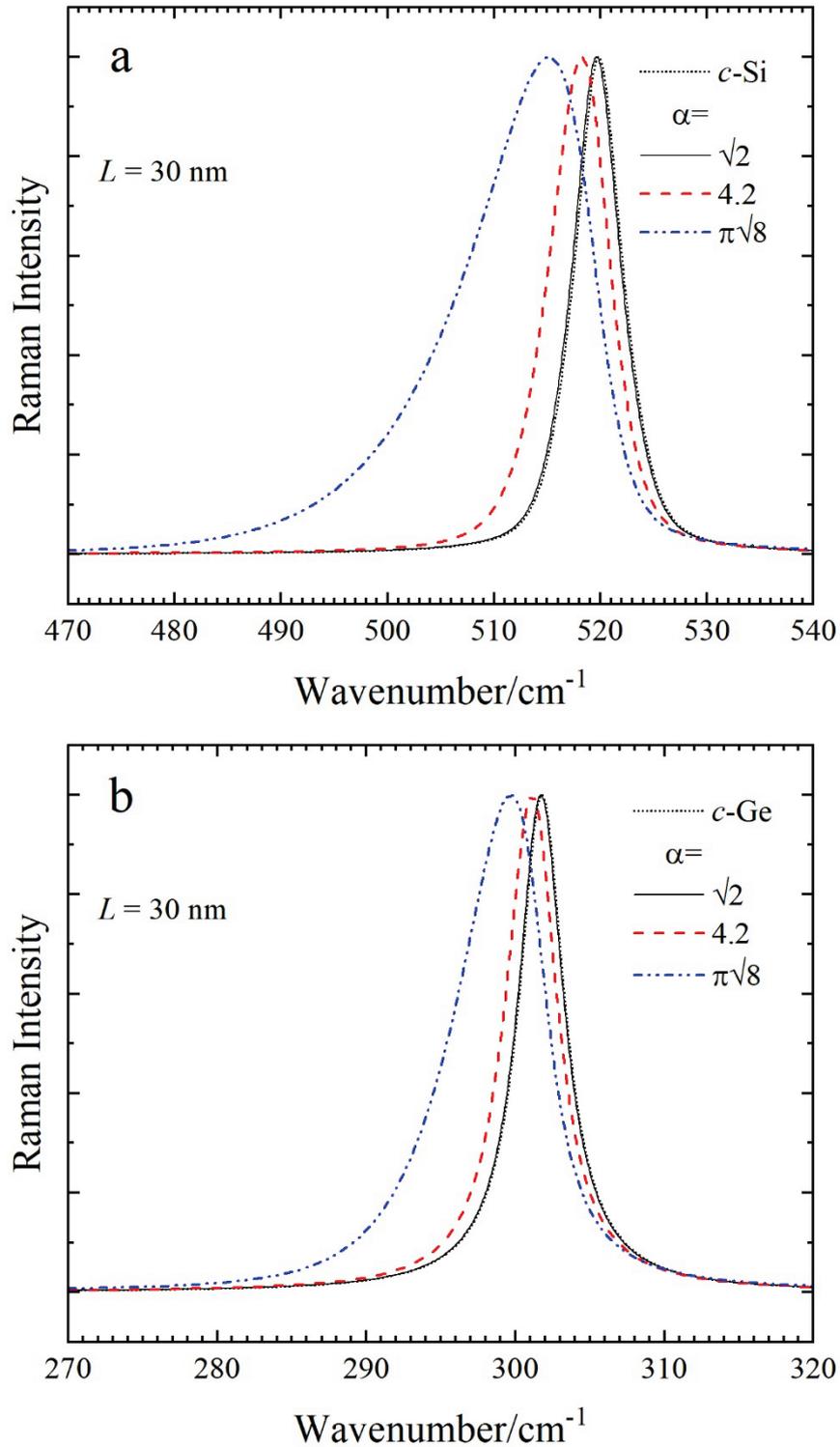

Fig. S2 Changes in the peak shape, position and width for different values of $\alpha$ in (a) Si and (b) Ge films; $L$ is an average size of crystal domains forming a polycrystalline film.

In Fig. S2, the simulated Raman spectra are presented depending on the value of the parameter $\alpha$ at fixed $L = 30$ nm. The spectra are simulated as the Lorentzian curves. There are noticeable differences in the downward shifts and asymmetrical broadening of the TO phonon band between Si and Ge samples. The changes are less pronounced in the Ge film than in the Si one.

Fig. S3 presents the Raman spectrum of the 50 nm thick Si layers grown at 500°C obtained using a 405-nm laser radiation. It is decomposed into three peaks: Peak 1, Peak 2 and Peak 3. The Peak 1 and Peak 2 components are combined into a single asymmetrical cumulative peak (Fig. S4). The simulated curve is a result of modeling of the Raman curve at $L=26$ nm. The calculated parameter $\alpha = 4.2$. For comparison, the c–Si peak obtained using the pseudo Voigt function for fitting is depicted in the position corresponding to the curve simulated at $L = 200$ nm.

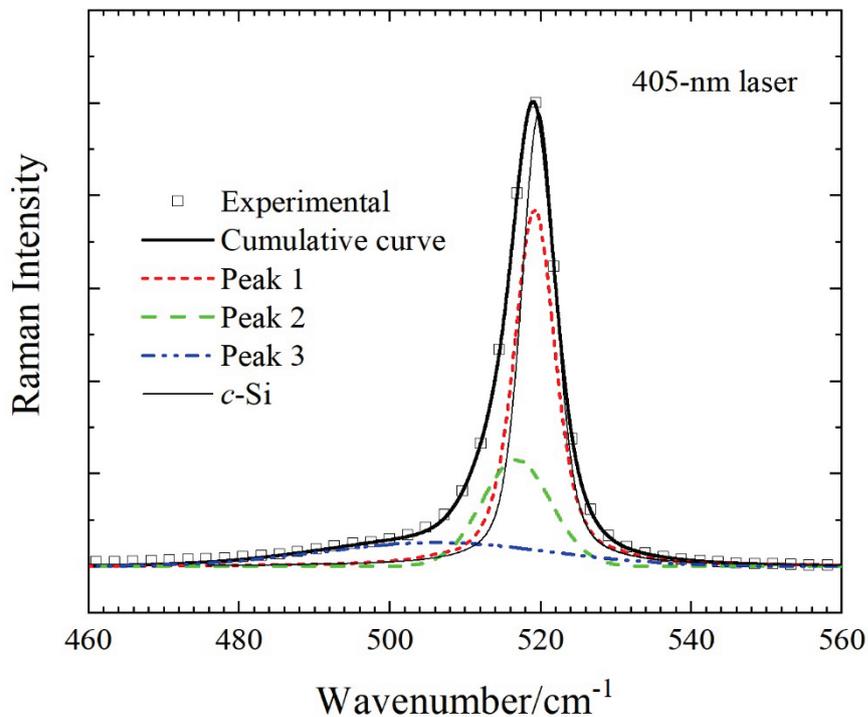

Fig. S3 Deconvolution of the Raman spectra of the 50-nm thick Si layers grown at 500°C (N3) in the vicinity of the TO(c-Si) vibration band peaked at 520 cm$^{-1}$; the c-Si spectrum is assigned to the Si$_3$N$_4$/SiO$_2$/Si (001) substrate; the laser wavelength is 405 nm; the spectrum is normalized to the maximum of the strongest peak at 520 cm$^{-1}$.

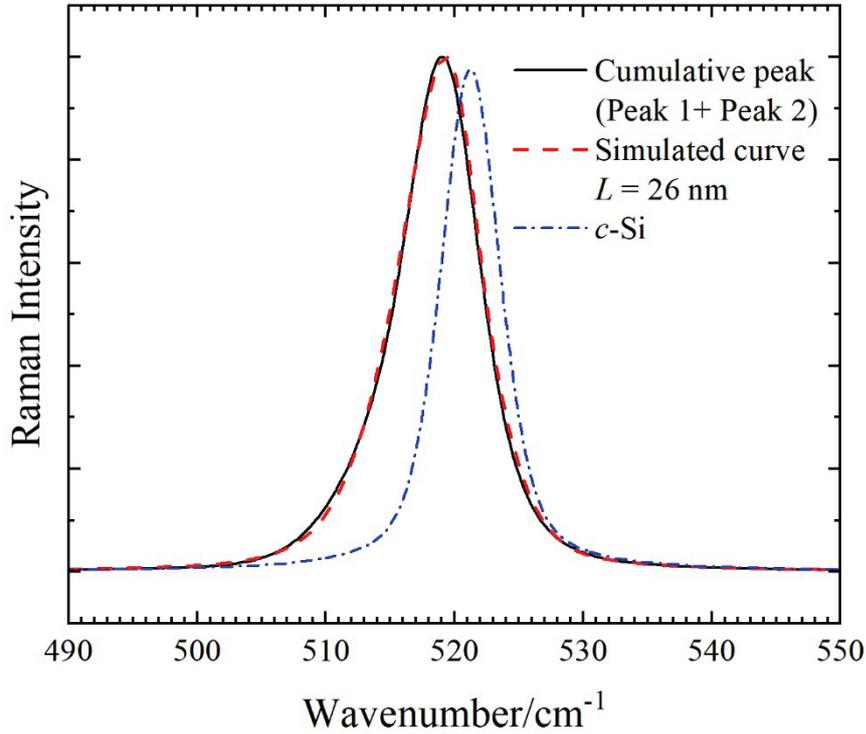

Fig. S4 Modeling of the Raman line shape of the peak composed of Peak 1 and Peak 2 components depicted in Fig. S3; the *c*-Si peak is that depicted in Fig. S1(b) (fitted using the pseudo Voigt function); its position corresponds to the curve simulated at $L$ = 200 nm.

Simulation of the phonon dispersion and the density of phonon states has been performed using the CASTEP Density functional modeling code (DFT) with GGA–PBE (Perdew–Burke–Ernzerhof) exchange correlation functional [1] and norm conserving pseudopotential. Full geometric optimization has been Full geometric optimization has been carried out using Quality Fine set: energy was less than $1\times10^{-5}$ eV/atom and the maximum force on any atom was less than 0.03 eV/Å). The Monkhorst-Pack grid parameters were set 3×3×2 for the reference sample and 2×2×1 for the structure with multiple twins. Cut off energy was 450 eV. Self-consistence field (SCF) tolerance was $1\times10^{-6}$ eV/atom (Fine).

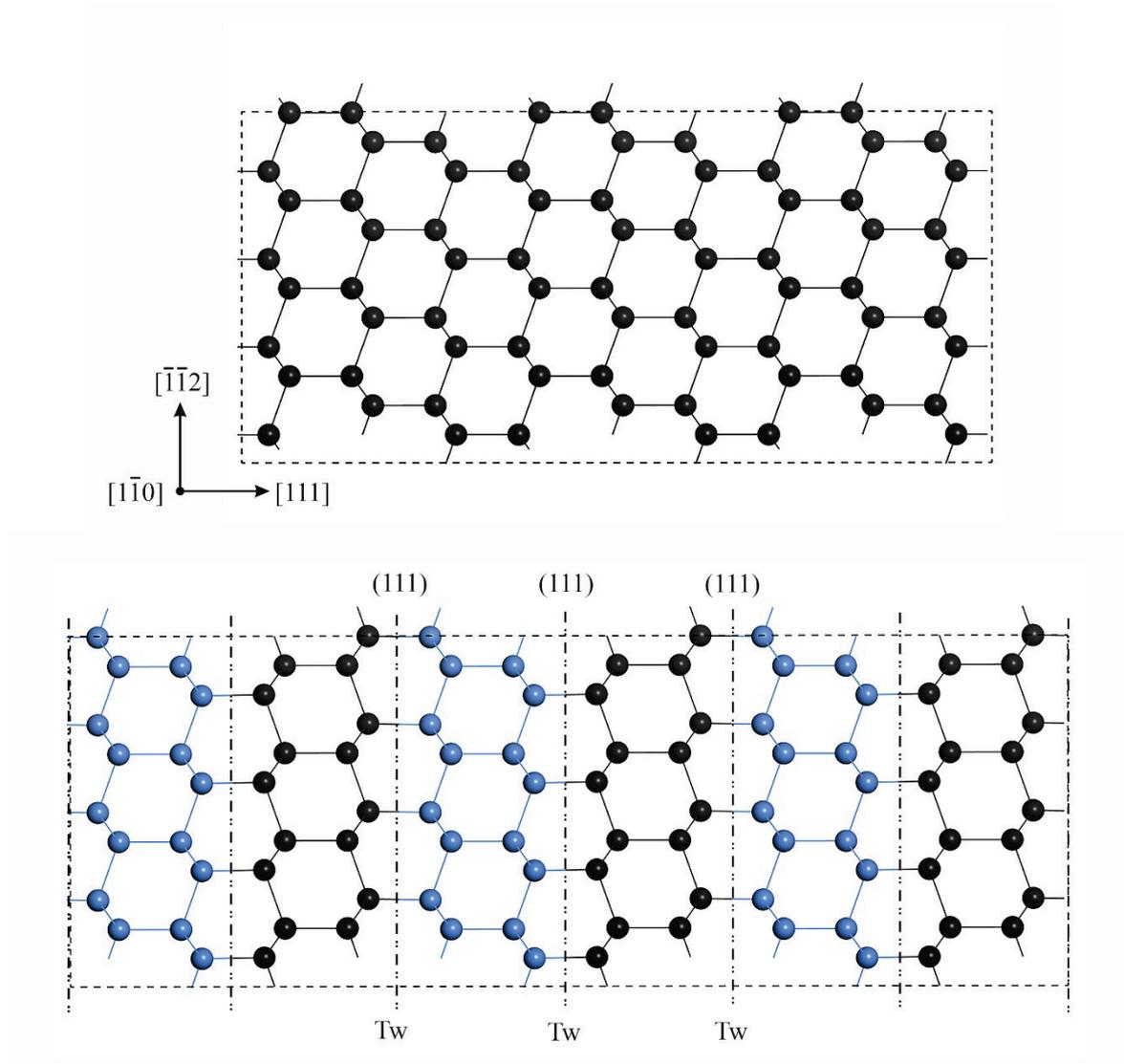

Fig. S5 Supercell models of the *c*-Si (*c*-Ge) perfect crystal (a) (reference sample) and multiple twin structure (b); in panel (b), *Tw* is the (111) plane and a twin boundary.

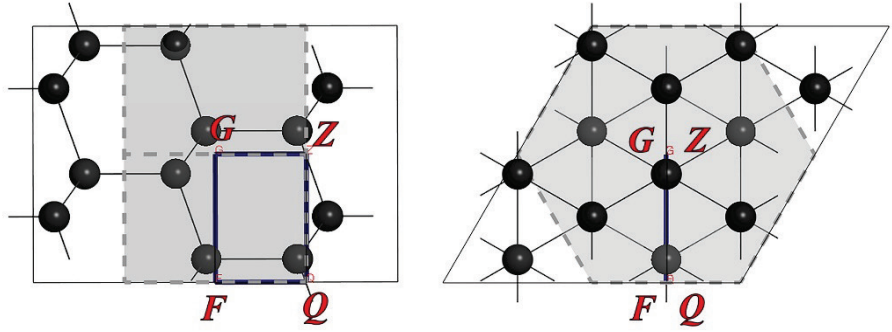

Fig. S6 The first Brillouin zone (highlighted in gray); the path along highly symmetrical directions G–F–Q–Z–G in reciprocal space is presented as a blue line.

For fully relaxed structures, we have computed the phonon dispersion and phonon DOS using a density-functional perturbation theory approach (linear response) [2] with 4×4×3 and 4×4×2 grids for the reference sample and the structure with multiple twins, respectively. Unit cells have been modelled from crystalline c-Si (c-Ge) with the diamond structure. The reference Si (Ge) sample was c-Si (c-Ge) unit cell cut by the (111) cleavage plane. The multiple twin structure consisted of a sequence of Σ3 [110] twins that were made up of alternating c-Si (c-Ge) unit cells of the (111) and ($\bar{1}\bar{1}1$) cleavage planes. They are presented in detail in Fig. S5.

Fig. S6 presents the first Brillouin zone of the modeled structures. The path is along high symmetrical directions G–F–Q–Z–G. Paths G–F and Q–Z are along <110> in the (111) twin boundary plane. F–Q and Z–G ones are along <111> directions.

Fig. S7 and Fig. S8 present results of simulation of the phonon dispersion and phonon DOS for the Si and Ge structures, respectively.

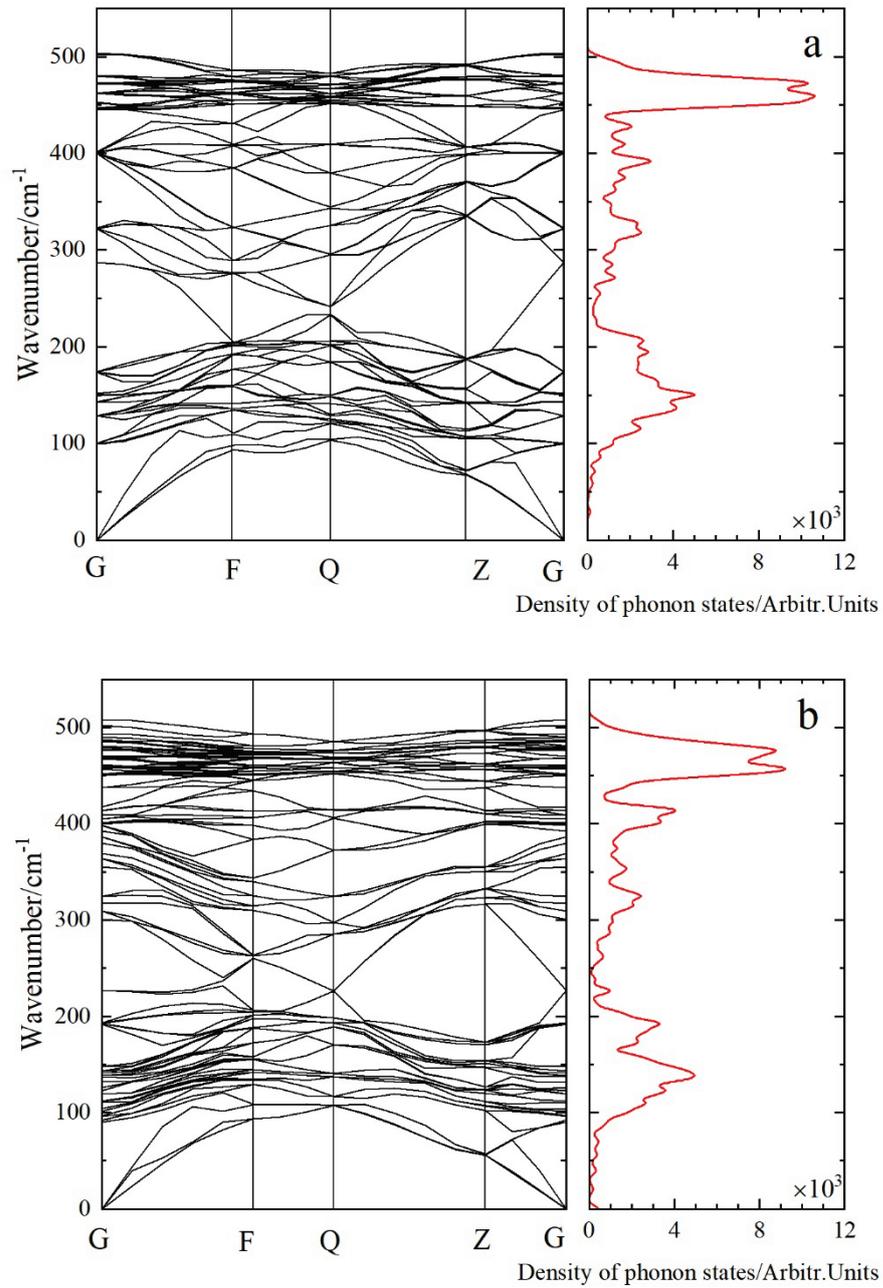

Fig. S7 Results of modeling of the phonon dispersion and phonon DOS for *c*-Si (a) and the multiple twin Si structure (b).

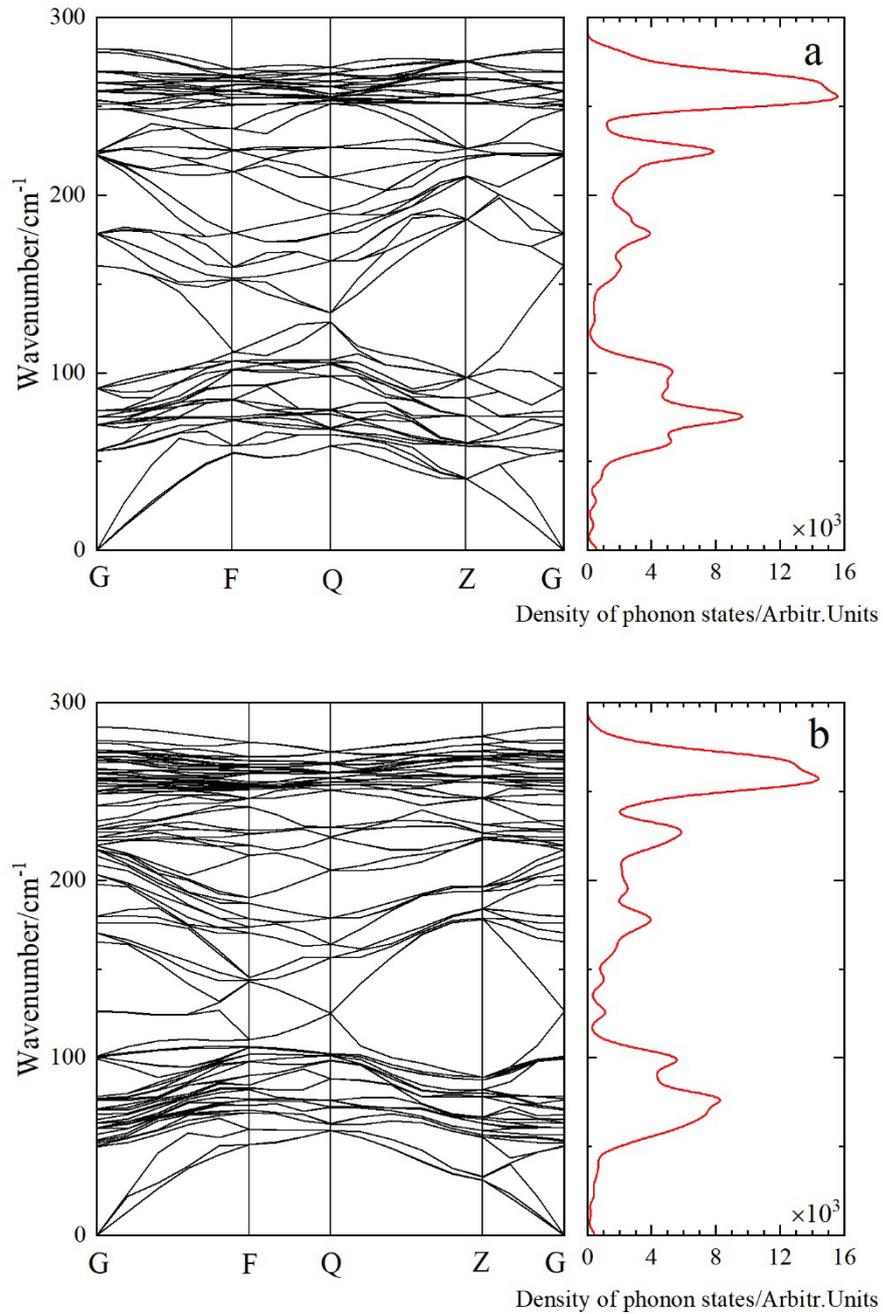

Fig. S8 Results of modeling of the phonon dispersion and phonon DOS for *c*-Ge (a) and the multiple twin Ge structure (b).

**References**

1. J.P. Perdew, K. Burke, M. Ernzerhof, *Phys. Rev. Lett.* **1996**, 77, 3865.
2. K. Refson, P.R. Tulip, S.J. Clark, *Phys. Rev. B* **2006**, 73, 155114.